\documentclass[12pt, twoside, here]{article}
\usepackage{epsf}
\usepackage{times,colordvi,amsmath,epsfig,float,color,multicol}
\usepackage{graphics}
\usepackage{hhline}
\usepackage[large]{subfigure}
\usepackage[latin1]{inputenc}
\usepackage{rotating}

\newtheorem{assumption}{\sc Assumption}

\title{A Model of Pupal Water Loss in {\em Glossina}}

\author{S. J. Childs \\ \\ {\small\em South African Centre for Epidemiological Modelling and Analysis,} \\ {\small\em University of Stellenbosch, Stellenbosch, 7600, South Africa}
\\ {\small\em tel: (+27/0) 21 8082589, email: schilds@sun.ac.za,}}

\renewcommand{\thefootnote}{\fnsymbol{footnote}}
\date{Mathematical Biosciences, 221: 77--90, 2009}       

\begin{document}

\maketitle
\renewcommand{\thefootnote}{\arabic{footnote}}

\begin{abstract}
\noindent {\em The results of a long-established investigation into pupal transpiration are used as a rudimentary data set. These data are then generalised to all temperatures and humidities by invoking the property of multiplicative separability, as well as by converting established relationships in terms of constant humidity at fixed temperature, to alternatives in terms of a calculated water loss. In this way a formulation which is a series of very simple, first order, ordinary differential equations is devised. The model is extended to include a variety of Glossina species using their relative surface areas, their relative pupal and puparial loss rates and their different 4th instar excretions. The resulting computational model calculates total, pupal water loss, consequent mortality and emergence. Remaining fat reserves are a more tenuous result. \\

\noindent The model suggests that, while conventional wisdom is often correct in
dismissing variability in transpiration-related pupal mortality as
insignificant, the effects of transpiration can be profound under adverse
conditions and for some species, in general. The model demonstrates how two
gender effects, the more significant one at the drier extremes of tsetse fly
habitat, might arise. The agreement between calculated and measured critical water losses suggests very little difference in the behaviour of the different species.
} 
\end{abstract}


\section{Introduction}

One of the more curious aspects of tsetse fly reproduction is that the larval
stages are initially retained in utero and become free living for only a very
short time. This is in contrast to many other insects which are more dependent
on water. Once the larva is deposited, it burrows into the substrate where it
excretes a protective case, known as a puparium. The puparium initially loses
water at a much higher rate than the pupa it encloses. The pupa receives no
external nourishment or fluids, whatsoever. Pupal development may be divided
into four successive stages viz. the third instar, the fourth instar, the sensu
strictu pupal stage and the pharate adult. Water loss rates adhere, in some
semblance, to this timetable and would appear to be little more than an
alternation between pupal and puparial rates. 

Early stage mortality is considered to be the most significant, by far, in any
model of tsetse population dynamics ({\sc Hargrove} \cite{Hargrove1}) and a
cursory inspection of the literature suggests pupal dehydration to be the most
challenging aspect of modelling it. The implications of pupal dehydration are
far greater than pupal emergence and mortality alone. Water loss continues after
eclosion up until the moment the teneral has its first meal. To give some idea
of relative importance, it can be argued that, while teneral water loss rates
are generally 20 times puparial rates and 100 times pupal rates (comparing {\sc
Bursell} \cite{Bursell1} with {\sc Bursell} \cite{Bursell2} data), puparial
rates generally prevail 6 times longer and pupal rates 24 times longer than
teneral rates. Thus, any teneral that dies of dehydration could be said to be,
at very least, 35\% as likely to have died as a result of pupal water loss. If
it is further specified that the teneral was of average age, the figure is
closer to 51\%. Water loss during the pupal phase can decide the fate of the
teneral. Combined dehydration and fat loss are thought to culminate in massive
teneral mortality. Pupal and teneral mortality rates are crucial in deciding the
viability of any tsetse population. The ultimate effect of cumulative water loss
on a given cohort is therefore likely to be best assessed in terms of the
proportion which have sufficient reserves to achieve their first feed. The
vastly different dynamics of water loss during the pupal and teneral phases,
however, afford pupal water loss the status of a topic in its own right. 

This work is almost exclusively based on the findings of one experimentalist
with all the hazards implied. In 1958, the late E. Bursell published the results
of his experiments on pupal water loss. Today, in an age of mainstream
computing, that work turns out to be a somewhat tantalizing, scientific
riddle. As one might imagine, Bursell's results are not of much use in the form
in which they were presented. Most of the work was carried out for steady
humidities at 24.7$^o$C. The main challenge to exploiting it for the purposes of a computational model, lies in generalising the results to all temperatures and humidities. That challenge could be said to be in three, very specific respects: A function for transpiration, the historical conditioning of sensu strictu pupal transpiration, then linking water loss to observed pupal emergence. Two further obstacles arise in the form of, firstly, extending the {\em Glossina morsitans}-based model to the rest of the {\em Glossina} genus and, secondly, resolving the dependence of emergence on humidity. 

The transpiration data were assigned to one of three categories for the purposes
of this work: Temperature dependent data, humidity dependent data and time
dependent data. One observes a certain amount of corroboration between points on
the respective curves. Some of this corroboration is demanded, for example, at
intersections, however, other of it comes as a pleasant surprise. The time
dependent data is a case in point. Time dependent transpiration would appear to
be nothing more than an alternation between pupal and puparial rates. The final
formulation, hence solution to the problem, is predicated on five major
assumptions. In addition to the assumptions explicitly stated and explored, two
others are taken for granted. The first is that it is assumed that {\sc Bursell}
\cite{Bursell1} is comprehensive, to the extent that it encapsulates all salient
aspects of pupal water loss. The second is that there is no transpirational
water loss at dewpoint. The problem is then reduced to a series of first order,
ordinary differential equations for water loss.

Although these equations are extremely simple, they are both numerous,
voluminous and there are issues pertaining to differentiability and continuity.
The relevant domains of applicability are also temperature dependent. This
renders preferred integration schemes, such as the fourth order accurate
Runge-Kutta-Fehlberg (R-K-F-4-5) method slightly impractical. Since the
problem is not intractably large, expedience takes precedence over taste and the
more pedestrian Euler's method is the preferred integration technique. The
resulting problem becomes a computational one, rather than a mathematical one. A
least squares fit, Newton's method and an half-interval search are the only
other techniques employed, the mathematics in the model being nothing more than
utility. 

This model is an experimental model. ``Experimental'' in the sense that it is
based on data gleened mostly from pre-1960s-published graphs and it relies
almost exclusively on the work of one experimentalist. It assumes that issues
such as inferior quality pupae, differing puparial durations and the shortage of
statistically significant data can be rectified at a later stage. Extending a
{\em G. morsitans}-based model to other species is work that can best be
described as exploratory. The results at the end make it an interesting and
justifiable exercise, nonetheless. 

\section{Generalising Scant Transpiration Data to a Function of Variable Humidity and Temperature}

{\sc Bursell} \cite{Bursell1} obtained one set of data points for variable
temperature (at 0\% r.h.), one for variable humidity (at 24.7 $\pm$ 3$^o$C) and
another for temporal dependence (at 0\% r.h. and 24.7 $\pm$ 3$^o$C), during his
investigations into pupal water loss. Yet a fourth set of data points can be
inferred by reason. One expects no transpirational water loss at dewpoint,
regardless of the temperature. Although of some assistance, the challenge,
nonetheless, remains: How does one generalise these data to all temperatures and
humidities? Fortunately, enough of the aforementioned data exist to suggest that
any transpiration function is not only continuous, it is also surprisingly
simple and smooth; monotonic, in fact.  
\begin{assumption} \label{assumption1} {\bf \em Transpiration rate is a multiplicatively, separable function of humidity and temperature. Put succinctly, if $\frac{dk}{dt}$ is the transpiration rate, then there exist two functions $\phi$ and $\theta$, dependent exclusively on humidity and temperature respectively, so that} 
\begin{eqnarray} \label{1}
\frac{dk}{dt}(h,T) &=& \phi(h) \ \theta(T),
\end{eqnarray} 
{\bf \em in which $h$ denotes humidity and $T$, the temperature.} 
\end{assumption} 
Just how reasonable is this assumption? Certainly it is consistent with, and
replicates, the fourth, inferred set of data points entertained above. The
perceived wisdom is that the region of interest is between 16$^o$C and 32$^o$C
(due, in part, to other causes of mortality). For the ``H'' of data which exists
across the humidity-temperature domain one reasonably expects rates to be
bounded by the wet-end and dry-end data, furthermore, to be close to
monotonic. One would also expect any unusual, capricious behaviour, or even
failure in the waterproofing, to manifest itself in dry air. The dry-air,
temperature-dependent data set is, fortunately, reasonably complete and
suggestive of behaviour which is smooth, monotonic and simple (pure exponential,
in this case). Thus, in the very likely event that water loss rates are not
multiplicatively separable, multiplicative seperability should not be a bad
substitute.

Of course, one can never be sure in these matters. Anything is possible. As much
as someone who models with data in the ideal format of a grid is ultimately
ignorant of the behaviour between grid points, one is faced here with the same
possibility of some magic combination of humidity and temperature. One simply
doesn't know; one can only surmise. Engineers make the same assumption under
what are sometimes, seemingly, a lot less favourable circumstances. Their
justification? It works. 


\subsection{The Dependence of Transpiration on Temperature}

Two types of transpirational temperature dependence are recognised in keeping with the multiplicative separability assumption. The first is a puparial type and the second is a pupal type. Both are exponential in nature. 

\subsubsection{For the Puparium}

Certainly so far as transpiration rates are concerned, for the puparium it is a
case of the worst first. Puparial transpiration rates oscillate wildly over time
and a fairly substantial difference in data which should corroborate is
documented. An average of the relevant data points from {\sc Bursell}
\cite{Bursell1} Figures 2a, 2b, 3 and 8a (those at 0\% r.h. and $24.7^0$C) were
accordingly used to adjust an exponential fit to the Figure 8a data upward. The
basis for this decision was threefold. Firstly, the Figure 8a data were read
from a logarithmic curve with a consequently greater, implied possibility of
error. Secondly, the humidity data were more comprehensively presented. Thirdly,
the study was overridingly an in-depth study of the effects of humidity,
strongly suggestive of an overall greater attention to detail and accuracy
pertaining to humidity dependence. The following function was the result,
\begin{eqnarray} \label{2}
\theta_{\mbox{\scriptsize puparium}}(T) &=& e^{0.110268 T - 9.92201} + 0.000354783,
\end{eqnarray} 
the units of which are {\em G. morsitans}, initial pupal masses per hour. The assymptotic standard errors\footnotemark[1] in fitting the constants for the exponential power were 6.8\% and 3.32\% respectively. The sum of squares of residuals\footnotemark[1] was 2.98448$\times 10^{-7}$. \footnotetext[1]{It is for consistency with this information that no attempt to guess the number of significant figures has been made.}
\begin{figure}[H]
    \begin{center}
\includegraphics[height=16cm, angle=-90, clip = true]{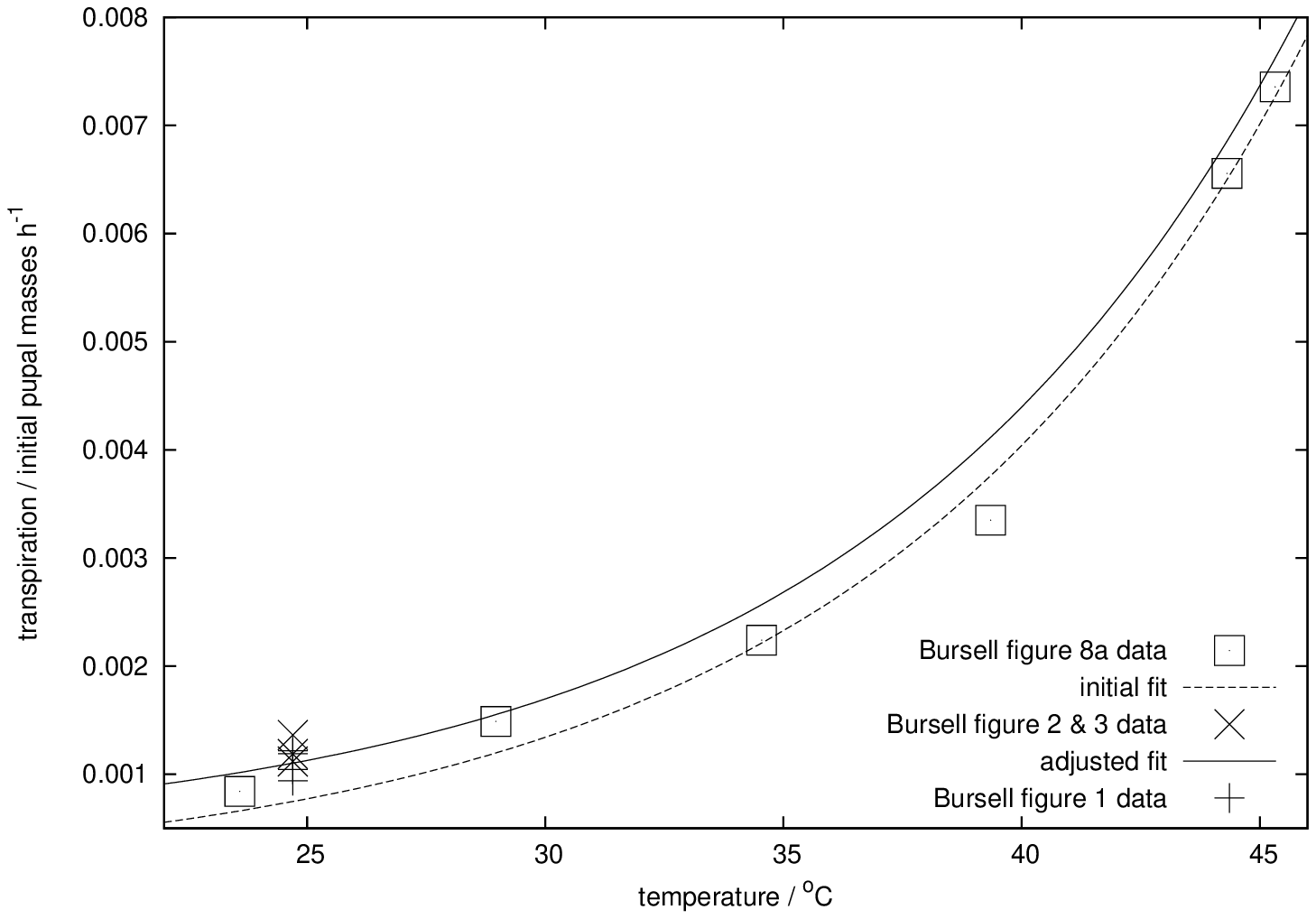}
\caption{The fit to the Bursell Figure 8a data was adjusted upward.} \label{puparialDependenceOnT}
   \end{center}
\end{figure} 
Of course, one might just as well have used the aforementioned graphs to argue the case for an upward adjustment of 0.00039619. There are many alternatives, however, a decision had to be made and so `modeller's licence' was invoked to make the choice which best takes cognizance of the, as yet unused, {\sc Bursell} \cite{Bursell1} Figure 1 data.

\subsubsection{For the Pupa}

Relevant data points from {\sc Bursell} \cite{Bursell1} Figures 5b and 6 (those at 0\% r.h. and $24.7^0$C) were added to the Figure 8b data of the same author. The following fit was obtained,
\begin{eqnarray} \label{3}
\theta_{\mbox{\scriptsize pupa}}(T) &=& e^{0.161691 T - 12.9591},
\end{eqnarray} 
the units of which are {\em G. morsitans}, initial pupal masses per hour. The assymptotic standard errors\footnotemark[1] in fitting the above constants were 4.597\% and 2.651\% respectively. The sum of squares of residuals\footnotemark[1] was 3.62324$\times 10^{-8}$. 

Note that, strictly speaking, the {\sc Bursell} \cite{Bursell1} Figure 8a data only pertains to the first day of the puparial duration. This will be of relevance in devising a temporal dependence.
\begin{figure}[H]
    \begin{center}
\includegraphics[height=16cm, angle=-90, clip = true]{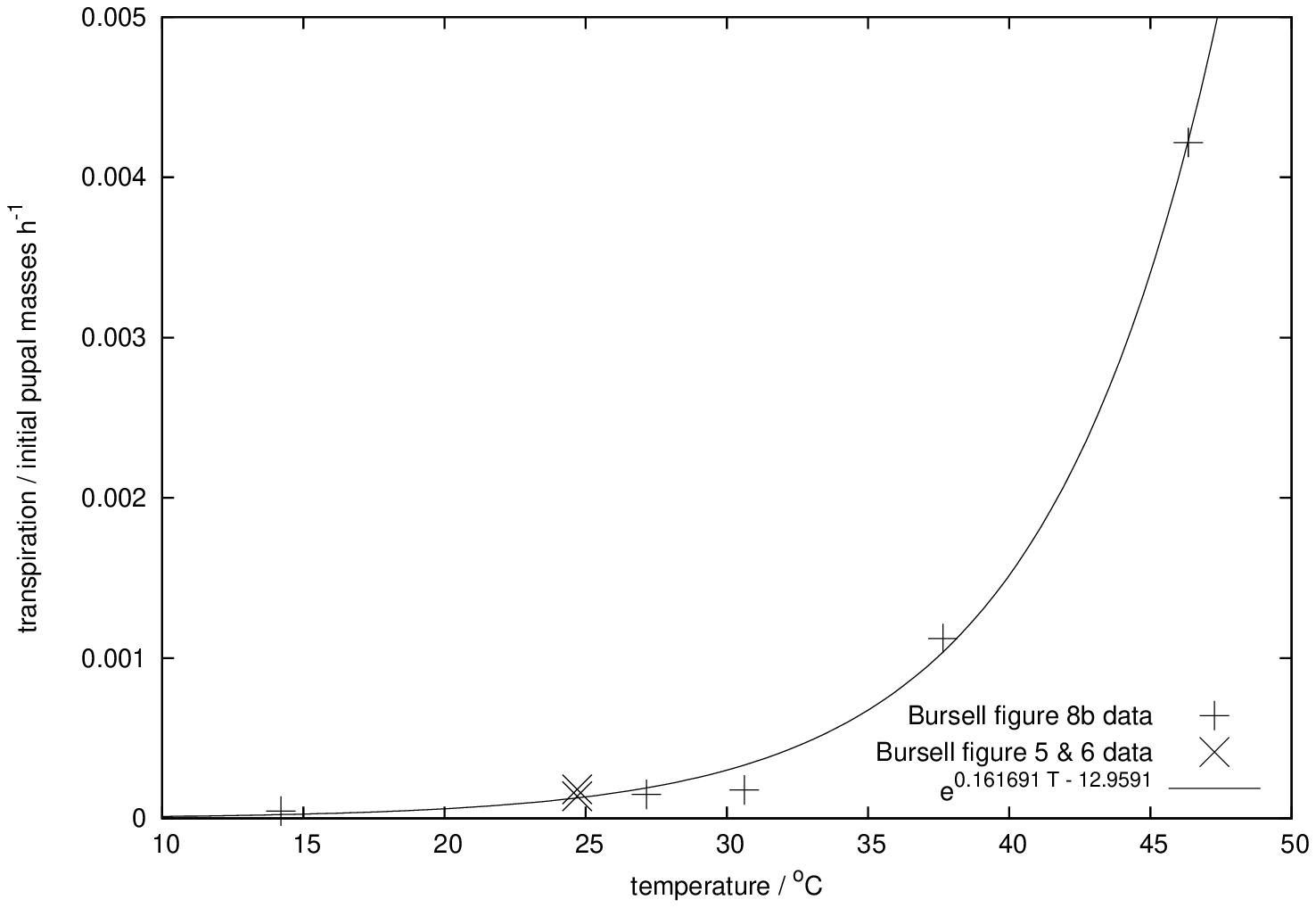}
\caption{The fit to the Bursell Figures 8b, 5 and 6 data.} \label{pupalDependenceOnT}
   \end{center}
\end{figure} 
			
\subsection{The Dependence of Transpiration on Humidity}

Two basic types of transpirational humidity dependence are recognised in keeping with the multiplicative separability assumption. The first is a puparial type and the second is a pupal type. Both are dimensionless.

\subsubsection{For the Puparium}

A brief inspection of {\sc Bursell} \cite{Bursell1} Figures 2a and 2b leads to the deduction that
\begin{eqnarray} \label{4}
\phi_{\mbox{\scriptsize puparium}}(h) &=& \frac{100 - h}{100}.
\end{eqnarray} 
The dependence of transpiration on humidity is linear.

\subsubsection{For the Pupa} \label{conditioning}

Pupal transpiration is not as straightforward. While the above relation may prevail during the initial transition down to pupal rates, one dependent on both total historical water loss and humidity is ultimately required. Historical conditioning, what one might term `drought hardening', alternatively depletion, is a phenomenon which pertains to tsetse pupae. During the sensu strictu pupal phase, transpiration becomes conditioned by the temperature and humidity which prevailed during the early stages (3rd and 4th instars inclusive). Present transpiration is conditioned by the recent past, in addition to the prevailing humidity and temperature. 

Relevant data is that published in Figures 5a, 5b and 6 of {\sc Bursell} \cite{Bursell1}. While it was, no doubt, acquired with a different purpose to the present one in mind and certainly proves a point, it is, of little use as presented. The problem is that historical humidities are steady, furthermore, the data were obtained at 24.7$^o$C. 
\begin{assumption} \label{assumption2} 
{\bf \em The transpiration rate, conditioned by a given historical water loss, is the same as the transpiration rate conditioned by an historically steady humidity, at 24.7$^{\mbox{o}}$C, which produced an equivalent total water loss.} \end{assumption} 
In other words, a conversion of the independent variable, historically-steady-humidity-at-24.7$^o$C, to an associated total water loss is inferred. Transpiration can then be re-expressed as a function of historical water loss. An historical conditioning of the pupa which is dependent on historically-non-steady variables is devised in this way. How reasonable is the assumption? Do different histories in temperature and humidity, which produce the same water loss, imply the same historical conditioning? If not, there are additional historical effects that have never been detected. 

The {\sc Bursell} \cite{Bursell1} Figures 5a, 5b and 6 data can be interpretted as transections through a surface which intersect at their ends, once their dependence on humidity has been converted to one of total historical water loss. They suggest a very simple surface, one which appears to be of no higher order than bi-quadratic, by inspection. It was therefore decided to fit a bi-quadratic surface to the historically conditioned transpiration data using the method of least squares, `on the fly' so-to-speak, that is
\begin{eqnarray*}
{\frac{dk}{dt}}_{\mbox{\scriptsize pupa}}(w,h,24.7) &=& c_1 + c_2h + c_3w + c_4wh + c_5h^2 + c_6w^2, \nonumber
\end{eqnarray*} 
in which $w$ is the total historical water loss and the $c_i$ are the constants
of the fit. The results were pleasing in that the surfaces retained their
fundamental character, even for fictitious, negatively-large humidities, the
importance of which will become apparent further on. 
\begin{figure}[H]
    \begin{center}
\includegraphics[height=11cm, angle=0, clip = true]{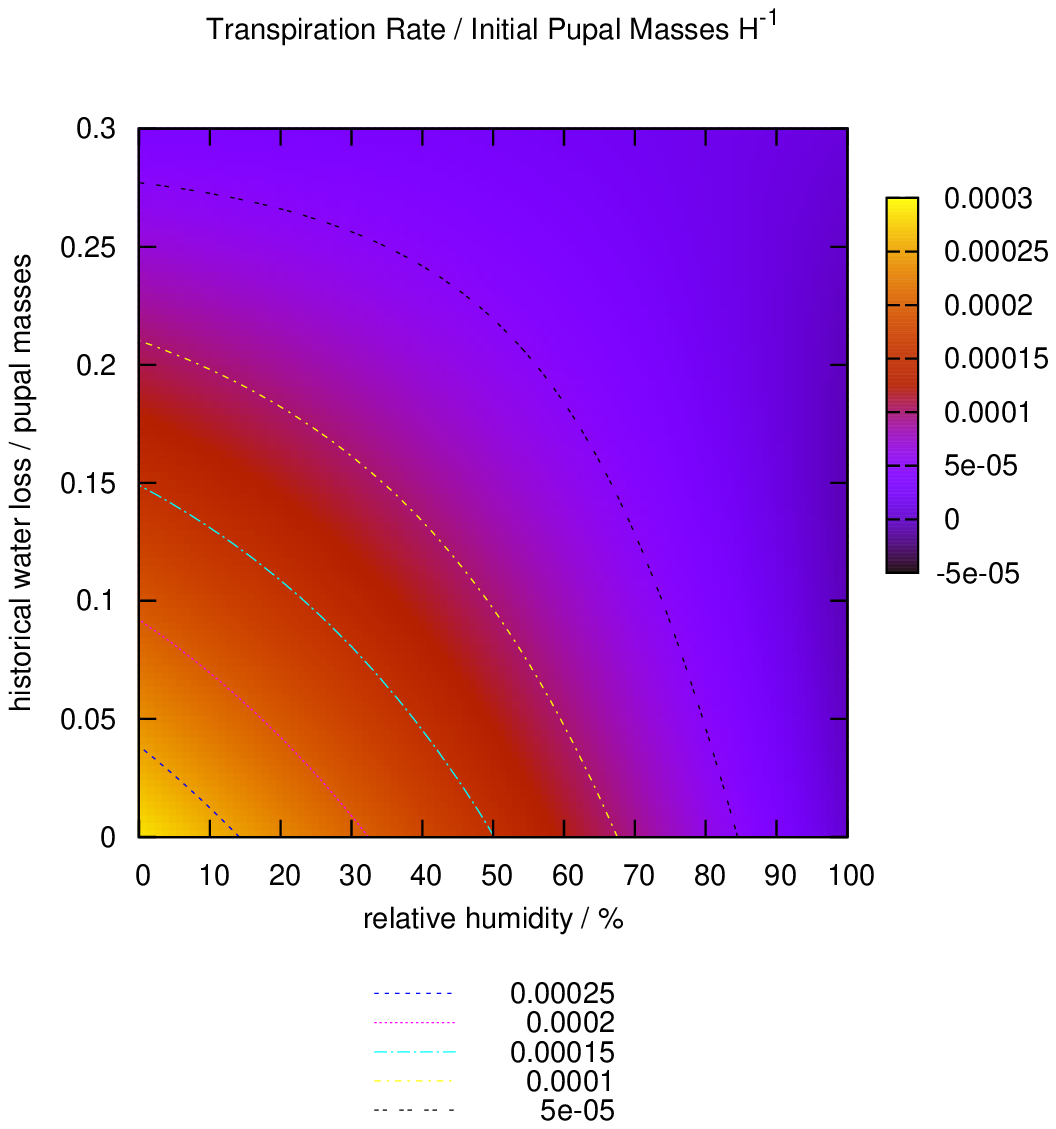}
\caption{The hourly transpiration rate as a function of humidity and water lost during the first 8/30 of the puparial duration (for {\em G. morsitans}).} \label{historicallyConditionedWaterLoss}
   \end{center}
\end{figure}
The surface is, nonetheless, a surface of transpiration rate when, instead, the
humidity dependence is sought. The following formulation
\begin{eqnarray} \label{6}
\phi_{\mbox{\scriptsize pupa}}(w,h) \ = \ \frac{{\frac{dk}{dt}}_{\mbox{\scriptsize pupa}}(w,h,24.7)}{\theta_{\mbox{\scriptsize pupa}}(24.7)} 
\end{eqnarray} 
can be derived based on the multiplicative separability assumption. Thus, the humidity dependence
\begin{eqnarray} \label{5}
\phi_{\mbox{\scriptsize pupa}}(w,h) &=& \frac{c_1 + c_2h + c_3w + c_4wh + c_5h^2 + c_6w^2}{\theta_{\mbox{\scriptsize pupa}}(24.7)}
\end{eqnarray} 
is obtained. 

\section{The Dependence of Transpiration on Time} \label{timeDependence}

Only the following stages of the puparial duration, $\tau$,
were deemed worthy of any time-dependent modelling, the vagaries of which were
entirely reduced to visually fitting four straight lines to the as-yet-unused
{\sc Bursell} \cite{Bursell1} Figure 1 data.

\subsubsection*{The Period $\mathbf{\frac{4}{30} \tau}$ to $\mathbf{\frac{6}{30} \tau}$}

During this period there is an adjustment from the puparial rate down to the pupal rate so that the total transpiration rate can be approximated as
\begin{eqnarray} \label{7}
\frac{dk}{dt} &=& {\frac{dk}{dt}}_{\mbox{\scriptsize puparium}} - \frac{27}{31}\frac{ \left( t - \frac{4}{30} \tau \right) }{ \left( \frac{6}{30} \tau - \frac{4}{30} \tau \right) }\left( {\frac{dk}{dt}}_{\mbox{\scriptsize puparium}} - {\frac{dk'}{dt}}_{\mbox{\scriptsize pupa}} \right), 
\end{eqnarray} 
in which $t$ is time, 
\begin{eqnarray*} 
{\frac{dk}{dt}}_{\mbox{\scriptsize puparium}} \ = \ \phi_{\mbox{\scriptsize puparium}}(h) \theta_{\mbox{\scriptsize puparium}}(T) \hspace{10mm} \mbox{and} \hspace{10mm} {\frac{dk'}{dt}}_{\mbox{\scriptsize pupa}} \ = \ \phi_{\mbox{\scriptsize puparium}}(h) \theta_{\mbox{\scriptsize pupa}}(T),
\end{eqnarray*} 
${\frac{dk'}{dt}}_{\mbox{\scriptsize pupa}}$ being a temporary, or transitional transpiration rate; one defined shortly prior to that for which the dependence on historical water loss is known.

\subsubsection*{The Period $\mathbf{\frac{6}{30} \tau}$ to $\mathbf{\frac{8}{30} \tau}$}

This period represents final adjustment down to the pupal rate. The equation
\begin{eqnarray} \label{8}
\frac{dk}{dt} &=& {\frac{dk}{dt}}_{\mbox{\scriptsize puparium}} - \frac{27}{31}\left( {\frac{dk}{dt}}_{\mbox{\scriptsize puparium}} - {\frac{dk'}{dt}}_{\mbox{\scriptsize pupa}} \right) - \frac{3}{31}\frac{ \left( t - \frac{6}{30} \tau \right) }{ \left( \frac{8}{30} \tau - \frac{6}{30} \tau \right) } \left( {\frac{dk}{dt}}_{\mbox{\scriptsize puparium}} - {\frac{dk'}{dt}}_{\mbox{\scriptsize pupa}} \right) \nonumber \\ &&
\end{eqnarray} 
was used.

\subsubsection*{The Period $\mathbf{\frac{8}{30} \tau}$ to $\mathbf{\frac{25}{30} \tau}$}

Transpiration during this phase is predominantly at the pupal rate. A small component of loss at puparial-rates increases linearly with time. The resulting combination was deemed to be
\begin{eqnarray} \label{9}
\frac{dk}{dt} &=& {\frac{dk}{dt}}_{\mbox{\scriptsize pupa}} + \frac{1}{30}\frac{ \left( t - \frac{8}{30} \tau \right) }{ \left( \frac{25}{30} \tau - \frac{8}{30} \tau \right) }\left( {\frac{dk}{dt}}_{\mbox{\scriptsize puparium}} - {\frac{dk}{dt}}_{\mbox{\scriptsize pupa}} \right), 
\end{eqnarray} 
in which
\begin{eqnarray*} 
{\frac{dk}{dt}}_{\mbox{\scriptsize pupa}} &=& \frac{\frac{dk}{dt}(w,h,24.7) \theta_{\mbox{\scriptsize pupa}}(T)}{\theta_{\mbox{\scriptsize pupa}}(24.7)}, 
\end{eqnarray*} 
based on equation \ref{6}.

\subsubsection*{The Period $\mathbf{\frac{25}{30} \tau}$ to $\mathbf{\frac{29}{30} \tau}$}

Transpiration begins its return to puparial rates during the pharate adult phase and is modelled by
\begin{eqnarray} \label{10}
\frac{dk}{dt} &=& {\frac{dk}{dt}}_{\mbox{\scriptsize pupa}} + \frac{1}{30}\left( {\frac{dk}{dt}}_{\mbox{\scriptsize puparium}} - {\frac{dk}{dt}}_{\mbox{\scriptsize pupa}} \right) + \frac{7}{30}\frac{ \left( t - \frac{25}{30} \tau \right) }{ \left( \frac{29}{30} \tau - \frac{25}{30} \tau \right) } \left( {\frac{dk}{dt}}_{\mbox{\scriptsize puparium}} - {\frac{dk}{dt}}_{\mbox{\scriptsize pupa}} \right) \nonumber \\ &&
\end{eqnarray} 
in this work.

\section{Extending the Model to Other Species}

It is generally suspected that the {\em Glossina} genus derives from a common, tropical, rain-forest dwelling ancestor, adjusted to moist, warm climates. One might therefore also suspect that all tsetse species actively pursue a strategy to minimise water loss for the majority of modern habitats and have hydrational mechanisms preventative of desiccation. It is generally accepted that most of the genus is not well adapted to arid environments {\sc Glasgow} \cite{Glasgow1}. The challenge to pupae, indeed the major threat, is dehydration. 

\begin{assumption} \label{assumption3} 
{\bf \em The hydrational mechanisms and water management strategies of the majority of tsetse fly species differ only with respect to relative pupal surface area, relative puparial loss rates, relative pupal loss rates, the different amounts excreted during the 4th instar and initial reserves.}
\end{assumption} 

Water loss rates for the puparium and pupa, $p_{\mbox{\scriptsize puparium}}$ and  $p_{\mbox{\scriptsize pupa}}$, respectively, have been measured for a number of species and are tabulated in {\sc Bursell} \cite{Bursell1}. Permeability is dependent on pressure and is quoted in units of $\mbox{mg \ h}^{-1}\mbox{cm}^{-2}(\mbox{mm Hg})^{-1}$. No variation with pronounced variation in temperature and humidity is indicated and it is of some comfort that the conversion of the {\em G. morsitans} model to other species involves relative rates.

Surface area data is likewise available. The same surface area is used for both the puparium and pupa in this work, the justification being that the puparial exuviae render the puparium marginally bigger while the pupal surface is not as regular. 

\subsection{For the Puparium}

A dimensionless, species conversion factor for puparial transpiration rates can be defined as follows
\begin{eqnarray} \label{11}
\delta_{\mbox{\scriptsize puparium}} &=& \frac{ p_{\mbox{\scriptsize puparium}} }{ p_{\mbox{\scriptsize morsitans puparium}} } \times \frac{ s_{\mbox{\scriptsize species}} }{ s_{\mbox{\scriptsize morsitans}} }, \nonumber 
\end{eqnarray}
in which $p_{\mbox{\scriptsize puparium}}$ is the rate of water loss for the species in question, $p_{\mbox{\scriptsize morsitans puparium}}$ is the equivalent water loss for {\em G. morsitans}, $s_{\mbox{\scriptsize species}}$ is the puparial surface area for the species in question and $s_{\mbox{\scriptsize morsitans}}$ is the equivalent, {\em G. morsitans} surface area. This factor enables the puparial transpiration rate for another species to be calculated from {\em G. morsitans} values. Note that the unit is still in {\em G. morsitans} initial pupal masses ($31 \mbox{mg}$) per hour. Actual values of $\delta_{\mbox{\scriptsize puparium}}$ for ten different species are tabulated in Table \ref{modelConversionFactors}.

\begin{table}[H]
    \begin{center}
\begin{tabular}{l l|c c c}  
&  &  &  & \\
Group & Species & $\delta_{\mbox{\scriptsize puparium}}$ & $\delta_{\mbox{\scriptsize pupa}}$ & $\delta_{\mbox{\scriptsize pupa}}$ \\ 
&  &  & (for minima) & (for maxima) \\
&  &  &  & \\ \hline 
&  &  &  & \\
{\em morsitans} & {\em austeni} \ & 1.60 & 0.712 & 0.723 \\ 
&  &  &  & \\ 
 \ & {\em morsitans} & 1 & 1 & 1 \\ 
&  &  &  & \\
 \ & {\em pallidipes} & 1.50 & 1.24 & 1.31 \\ 
&  &  &  & \\
 \ & {\em submorsitans} & 2.44 & 0.950 & --\\ 
&  &  &  & \\
 \ & {\em swynnertoni} & 0.830 & 0.869 & 0.892 \\ 
&  &  &  & \\
{\em palpalis} & {\em palpalis} & 2.54 & 1.41 & 1.36 \\ 
&  &  &  & \\ 
 \ & {\em tachinoides} & 0.818 & 0.743 & -- \\ 
&  &  &  & \\
{\em fusca} & {\em brevipalpis} & 10.3 & 4.57 & 3.06 \\ 
&  &  &  & \\ 
 \ & {\em fuscipleuris} & 8.84 & 4.45 & 3.16 \\ 
&  &  &  & \\
 \ & {\em longipennis} & 3.62 & 2.45 & 2.30 \\ 
&  &  &  & \\
\end{tabular}
\caption{Species conversion factors for the model calculated from data, ultimately sourced from {\sc Buxton} and {\sc Lewis} \cite{BuxtonAndLewis1}, presented in {\sc Bursell} \cite{Bursell1}.
} \label{modelConversionFactors}
    \end{center}
\end{table}

\subsection{For the Pupa}

A dimensionless, species conversion factor for pupal transpiration rates can be defined as follows
\begin{eqnarray} \label{12}
\delta_{\mbox{\scriptsize pupa}} &=& \frac{ p_{\mbox{\scriptsize pupa}} }{ p_{\mbox{\scriptsize morsitans pupa}} } \times \frac{ s_{\mbox{\scriptsize species}} }{ s_{\mbox{\scriptsize morsitans}} } \nonumber 
\end{eqnarray} 
in which $p_{\mbox{\scriptsize pupa}}$ is the rate of water loss for the species
in question, $p_{\mbox{\scriptsize morsitans pupa}}$ is the equivalent water
loss for {\em G. morsitans}, $s_{\mbox{\scriptsize species}}$ is the pupal
surface area for the species in question and $s_{\mbox{\scriptsize morsitans}}$
is the equivalent, {\em G. morsitans} surface area. This factor enables the
puparial transpiration rate for another species to be calculated from {\em G.
morsitans} values. Note that the unit is still in {\em G. morsitans} initial
pupal masses ($31 \mbox{mg}$) per hour. Actual values of
$\delta_{\mbox{\scriptsize pupa}}$ for ten different species are tabulated in
Table \ref{modelConversionFactors}.

On the face of it, Assumption \ref{assumption3} is certainly the most tenuous.
How valid is it? Does such a simplistic approach work? Very little data is
available for other species, however, 3rd instar, puparial loss rates for both 
{\em Glossina brevipalpis} and {\em Glossina palpalis} are known. Conversion of
3rd instar, G. morsitans-model, transpiration rate values to {\em G.
brevipalpis} and {\em G. palpalis} values, yielded errors of 6\% and 10\%
respectively. The $\delta_{\mbox{\scriptsize pupa}}$ for pupal maxima and minima
in Table \ref{modelConversionFactors} are, furthermore, remarkably similar (for
all except {\em G. brevipalpis} and {\em Glossina fuscipleuris}). This is very
encouraging and suggestive of a similar slope in the transpirational time
dependence for the various species. The suggestion for {\em G. brevipalpis} and
{\em G. fuscipleuris}, however, is that Assumption \ref{assumption3} could
possibly be captious.

What are the implications? It means that knowing only the appropriate {\em G. morsitans} $\phi$ and $\theta$ is adequate. The only questions pertaining to species conversion which remain are whether the temporal interplay between pupal and puparial transpiration rates is the same for the entire puparial duration and whether the historical conditioning is the same; not withstanding some difference in strategy a la the difference in 4th instar excretions\footnotemark[1].\footnotetext[1]{Although, in that case, the associated water loss is too small to be of any real consequence.} 

%
%
%

\section{The Resulting Model for Pupal Water Loss}

Taking into account one, further formula (that for puparial duration), results in a model.

\subsection{The Puparial Duration}

The puparial duration in days, $\tau$, is calculated according to the formula
\begin{eqnarray} \label{13}
\tau &=& \frac{ 1 + e^{a + bT} }{\kappa},
\end{eqnarray} 
in which $T$ is temperature ({\sc Phelps and Burrows} \cite{phelpsAndBurrows1},
modified by {\sc Hargrove} \cite{Hargrove3}). For females, $\kappa = 0.057 \pm
0.001$, $a = 5.5 \pm 0.2$ and $b = -0.25 \pm 0.01$. For males, $\kappa = 0.053
\pm 0.001$, $a = 5.3 \pm 0.2$ and $b = -0.24 \pm 0.01$. The puparial durations
of all species, with the exception of {\em G. brevipalpis}, are thought to lie
within 10\% of the value predicted by this formula {\sc Parker} \cite{Parker1}.
{\em G. brevipalpis} takes a little longer.

Newton's method is used to solve for a puparial duration based on the daily average, which is, of course, dependent on itself. The same applies for the various fractions of puparial duration. For this reason the notation $\tau_{\frac{r}{30}}$ is adopted, where
\begin{eqnarray*} 
\tau_{\frac{r}{30}} &\equiv& \frac{1}{\tau_{\frac{r}{30}}} \left[ \left( \tau_{\frac{r}{30}} - \mbox{floor}\left\{ \tau_{\frac{r}{30}} \right\} \right) \ \frac{r}{30} \ \tau(T_{{\mbox{\scriptsize day floor}} \left\{ \tau_{\frac{r}{30}} \right\} + 1}) \ + \sum_{i=1}^{{\mbox{\scriptsize floor}} \left\{ \tau_{\frac{r}{30}} \right\}} \frac{r}{30} \ \tau(T_{\mbox{\scriptsize day }{\scriptsize i}}) \right],
\end{eqnarray*} 
it being the average $\frac{r}{30} \times \tau$ over the time interval $(0, \tau_{\frac{r}{30}}]$, for a specified $r$.

\subsection{The Governing Equations}

Collecting together all prior observations and thoughts gives rise to the
following series of first order, ordinary differential equations. Note that what would otherwise have been a unit of $31 \mbox{mg h}^{-1}$ can be replaced by a dimensionless
\begin{eqnarray*}
&& \frac{ m_{\mbox{\scriptsize morsitans}} }{ m_{\mbox{\scriptsize species}} } \ \mbox{initial pupal masses} \times \frac{ 24 \tau_{1} }{ 1 \  \mbox{puparial duration} }
\end{eqnarray*} 
for the usual reasons, where $m_{\mbox{\scriptsize morsitans}}$ and $m_{\mbox{\scriptsize species}}$ are the initial pupal masses of {\em G. morsitans} (31mg) and the species in question respectively. In other words, a dimensionless rate unit of initial-pupal-masses per puparial-duration is preferred. 

\subsubsection*{The Period 0 to $\mathbf{\tau_{\frac{4}{30}}}$}

The water loss rate for the greater part of the third and fourth instars is at the puparial rate (obtained by the substitution of equation \ref{2} and equation \ref{4} into equation \ref{1}). Generalising the resulting expression to all species and writing the equation in dimensionless form results in
\begin{eqnarray} \label{14}
\frac{dk}{dt} &=& ( e^{0.110268 T - 9.92201}  + 0.000354783 ) \frac{100 - h}{100} \ \frac{ p_{\mbox{\scriptsize puparium}} }{ p_{\mbox{\scriptsize morsitans puparium}} } \frac{ s_{\mbox{\scriptsize species}} }{ s_{\mbox{\scriptsize morsitans}} }. 
\end{eqnarray}
\subsubsection*{The Period $\mathbf{\tau_{\frac{4}{30}}}$ to $\mathbf{\tau_{\frac{6}{30}}}$}

During this period there is an adjustment from the puparial rate down to the pupal rate dictated by equation \ref{7}. Generalising equation \ref{7} to all species and writing the equation in dimensionless form results in the expression
\begin{eqnarray} \label{15}
\frac{dk}{dt} &=& \left[ ( e^{0.110268 T - 9.92201}  + 0.000354783 ) \left( 1 - \frac{27}{31} \frac{ \left( t' - \tau_{\frac{4}{30}} \right) }{ \left( \tau_{\frac{6}{30}} - \tau_{\frac{4}{30}} \right) } \right) \right. \frac{ p_{\mbox{\scriptsize puparium}} }{ p_{\mbox{\scriptsize morsitans puparium}} } \nonumber \\ 
&& + \ \left. e^{0.161691 T - 12.9591} \frac{27}{31}\frac{ \left( t' - \tau_{\frac{4}{30}} \right) }{ \left( \tau_{\frac{6}{30}} - \tau_{\frac{4}{30}} \right) } \ \frac{ p_{\mbox{\scriptsize pupa}} }{ p_{\mbox{\scriptsize morsitans pupa}} } \right] \frac{100 - h}{100} \frac{ s_{\mbox{\scriptsize species}} }{ s_{\mbox{\scriptsize morsitans}} },
\end{eqnarray}
in which $t'$ is a developmental `time', $t' = \tau_{\frac{t}{\tau_{1}}}$.

\subsubsection*{The Period $\mathbf{\tau_{\frac{6}{30}}}$ to $\mathbf{\tau_{\frac{8}{30}}}$}

During this period there is a final adjustment from the puparial rate down to the pupal rate dictated by equation \ref{8}. Generalising equation \ref{8} to all species and writing the equation in dimensionless form results in
\begin{eqnarray} \label{16}
\frac{dk}{dt} \ = \ \left[ ( e^{0.110268 T - 9.92201}  + 0.000354783 ) \left( \frac{4}{31} - \frac{3}{31}\frac{ \left( t' - \tau_{\frac{6}{30}} \right) }{ \left( \tau_{\frac{8}{30}} - \tau_{\frac{6}{30}} \right) } \right) \right.\frac{ p_{\mbox{\scriptsize puparium}} }{ p_{\mbox{\scriptsize morsitans puparium}} } && \nonumber \\ 
+ \ \left. e^{0.161691 T - 12.9591} \left( \frac{27}{31} + \frac{3}{31}\frac{ \left( t' - \tau_{\frac{6}{30}} \right) }{ \left( \tau_{\frac{8}{30}} - \tau_{\frac{6}{30}} \right) } \right) \frac{ p_{\mbox{\scriptsize pupa}} }{ p_{\mbox{\scriptsize morsitans pupa}} } \right] \frac{100 - h}{100} \frac{ s_{\mbox{\scriptsize species}} }{ s_{\mbox{\scriptsize morsitans}} }. &&  
\end{eqnarray}
\subsubsection*{Excretion} 
If water loss is sufficiently low during the first $\frac{8}{30}$ of the puparial duration, cognizance must be taken of the small amount excreted. In this unlikely scenario the formula
\begin{eqnarray*} 
k &=& x_{\mbox{\scriptsize 2}} + \frac{h_{\mbox{\scriptsize 3rd instar}} }{ 100 } ( x_{\mbox{\scriptsize 1}} - x_{\mbox{\scriptsize 2}} ) 
\end{eqnarray*} 
was implemented, for want of any better wisdom. The total water loss during the 3rd and 4th instars, in the event of dewpoint prevailing for the former, is $x_{\mbox{\scriptsize 1}}$. In the event of 0\% relative humidity prevailing for the 3rd instar, the amount is $x_{\mbox{\scriptsize 2}}$. The only 4th instar excretion data known to exist is that for {\em G. morsitans}, {\em G. palpalis} and {\em G. brevipalpis}. This lack of information is a minor obstacle as the excretions are generally small and only relevant for humidities close to dewpoint.
 
\subsubsection*{The Period $\mathbf{\tau_{\frac{8}{30}}}$ to $\mathbf{\tau_{\frac{25}{30}}}$}

Transpiration during this phase is predominantly at the pupal rate. There is also deemed to be a small component of loss at puparial-rates, which increases linearly with time and which is included for good measure. Generalising equation \ref{9} to all species and writing the equation in dimensionless form results in
\begin{eqnarray} \label{17}
\frac{dk}{dt} &=& \left[ e^{0.161691 ( T - 24.7 )} (c_1 + c_2h + c_3w + c_4wh + c_5h^2 + c_6w^2) \left(1 - \frac{1}{30}\frac{ \left( t' - \tau_{\frac{8}{30}} \right) }{ \left( \tau_{\frac{25}{30}} - \tau_{\frac{8}{30}} \right) }\right) \right. \nonumber \\ 
&& \times \frac{ p_{\mbox{\scriptsize pupa}} }{ p_{\mbox{\scriptsize morsitans pupa}} } \ + \ ( e^{0.110268 T - 9.92201}  + 0.000354783 ) \ \frac{100 - h}{100} \nonumber \\ 
&& \left. \times \frac{1}{30} \ \frac{ \left( t' - \tau_{\frac{8}{30}} \right) }{ \left( \tau_{\frac{25}{30}} - \tau_{\frac{8}{30}} \right) } \ \frac{ p_{\mbox{\scriptsize puparium}} }{ p_{\mbox{\scriptsize morsitans puparium}} } \right] \frac{ s_{\mbox{\scriptsize species}} }{ s_{\mbox{\scriptsize morsitans}} }.
\end{eqnarray}
\subsubsection*{The Period $\mathbf{\tau_{\frac{25}{30}}}$ to $\mathbf{\tau_{\frac{29}{30}}}$}

Transpiration begins its return to puparial rates during the pharate adult phase. Generalising equation \ref{10} to all species and writing the equation in dimensionless form results in
\begin{eqnarray} \label{18}
\frac{dk}{dt} &=& \left[ e^{0.161691 ( T - 24.7 )} (c_1 + c_2h + c_3w + c_4wh + c_5h^2 + c_6w^2) \left( \frac{29}{30} - \frac{7}{30}\frac{ \left( t' - \tau_{\frac{25}{30}} \right) }{ \left( \tau_{\frac{29}{30}} - \tau_{\frac{25}{30}} \right) } \right) \right. \nonumber \\ 
&& \times \frac{ p_{\mbox{\scriptsize pupa}} }{ p_{\mbox{\scriptsize morsitans pupa}} } \ + \ ( e^{0.110268 T - 9.92201}  + 0.000354783 ) \ \frac{100 - h}{100}  \nonumber \\ 
&& \left. \times \left( \frac{1}{30} + \frac{7}{30}\frac{ \left( t' - \tau_{\frac{25}{30}} \right) }{ \left( \tau_{\frac{29}{30}} - \tau_{\frac{25}{30}} \right) } \right) \ \frac{ p_{\mbox{\scriptsize puparium}} }{ p_{\mbox{\scriptsize morsitans puparium}} } \right] \frac{ s_{\mbox{\scriptsize species}} }{ s_{\mbox{\scriptsize morsitans}} }. 
\end{eqnarray}
\subsubsection*{The Period $\mathbf{\tau_{\frac{29}{30}}}$ to $\mathbf{\tau_1}$}

There is a return to puparial rates shortly before eclosion and equation \ref{14} once again applies. 

\subsection{Solving the Equations}

The above rate formulae constitute a series of first order, ordinary
differential equations. One expects the resulting function to be Lipshitz
continuous over each of the developmental sub-stages identified, likely even a
contraction. While a fourth-order-accurate Runge-Kutta-Fehlberg
(R-K-F-4-5) method\footnotemark[1] \footnotetext[1]{See any standard
textbook on numerical analysis for further information.} would normally be the
preferred method of integration, the series of equations is voluminous, issues
of non-differentiabilty and discontinuity pertain to the complete pupal period
and the relevant subdomains of applicability are both numerous and temperature
dependent. 

Euler's method is usually considered distasteful from the point of view of its
error. The local error per step, of length ${\Delta t}$, is $O({\Delta t}^2)$.
Since the required number of steps is proportional to $\frac{1}{{\Delta t}}$,
the global error is $O({\Delta t})$. This is indeed primitive. The method is,
nonetheless, considered robust for the type of first order, ordinary
differential equation to be solved. The real strength of Euler's method lies in
its robustness at discontinuities and points of non-differentiability. The
maximum, additional error introduced at such points is of the same order as the
method's global error (this is easy to see). The same cannot be said for the
higher order methods. The use of one or other of the higher order methods is
still not precluded in the problem at hand, since the discontinuities and points
at which differentiability breaks down are predictable. Using Euler's method,
however, one has one problem to solve, whereas using one of the higher order
methods entails solving six, seperate problems; each confined to its own
respective domain of Lipshitz continuity, requiring dynamic scaling etc.. 

The handicap of a poor error is easily overcome computationally. That is by using a small step length e.g. $\frac{1}{1000}$th of a puparial duration. 
Two significant figures are all that are sought. Since the problem is not
intractably large, expedience takes precedence over taste and the more
pedestrian Euler's method is considered the appropriate choice. 

\section{Pupal Emergence and Mortality} \label{emergence}

Two challenges arise when it comes to pupal emergence: The first is to establish some kind of credible relationship between the numbers of emergent and humidity. The second is, consequently, how to relate emergence to total water loss.

\subsection{What is the Relationship Between Pupal Emergence and Humidity?} \label{gaussian}

\begin{figure}[H]
    \begin{center}
\includegraphics[height=16cm, angle=-90, clip = true]{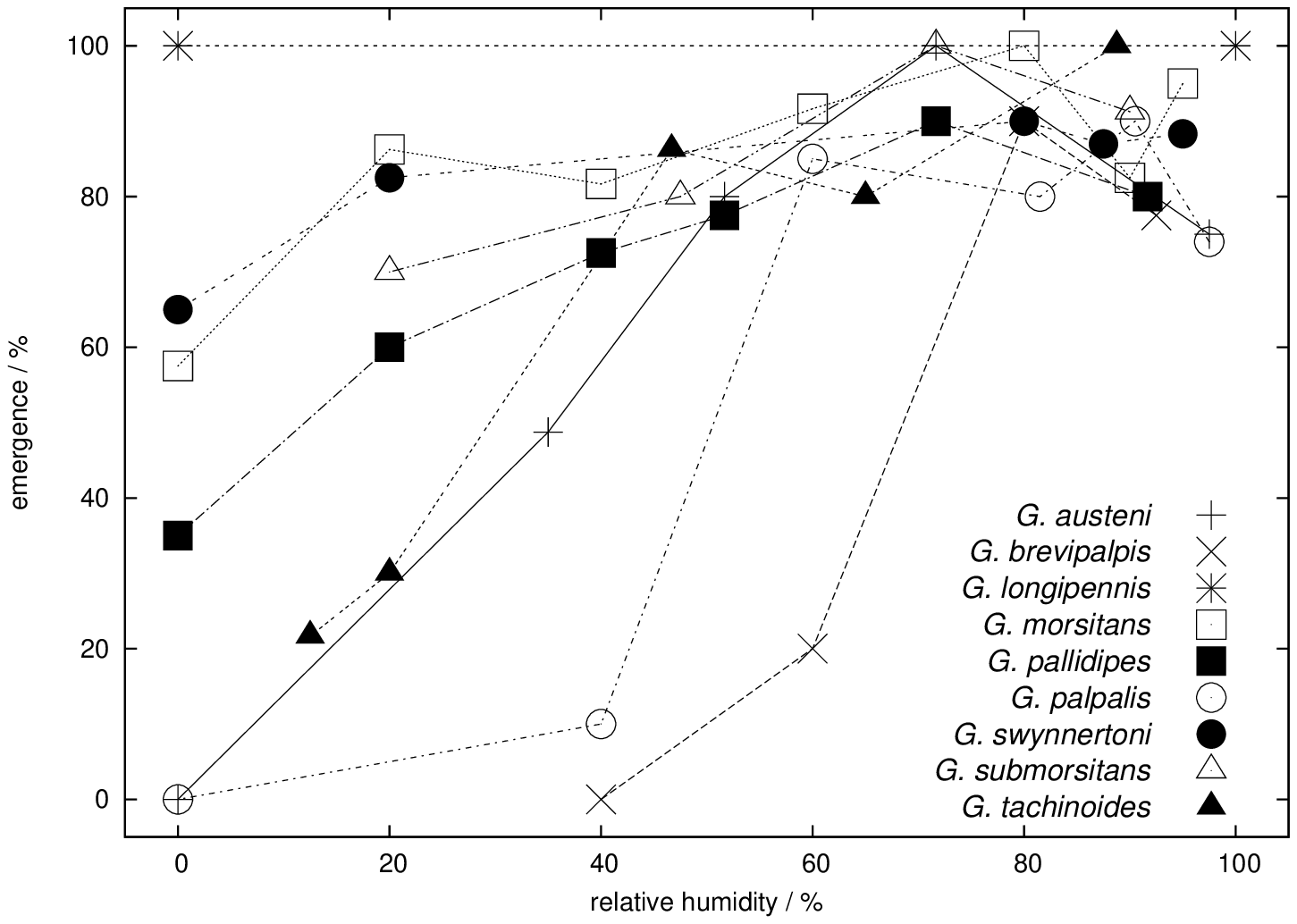}
\caption{Emergence data as presented by {\sc Bursell} \cite{Bursell1}. All are at 24$^o$C, except {\em G. tachinoides} (30$^o$C).} \label{emergenceData}
   \end{center}
\end{figure}

What does one make of the very rudimentary data in Figure \ref{emergenceData}? 
There are insufficient data points for each species to perform any kind of
rigorous hypothesis testing.  Since the number of pupae contributing to each
data point is so low, neither the law of large numbers can be invoked, nor
therefore, the central limit theorem applied. This does, nonetheless, not
necessarily preclude the use of the aforementioned in argument. Large numbers of
pupae do exist. When doctrinaire methods fail a little thought can still go a
long way.

At the simplest level, one would expect each species to be adapted to some ideal humidity for which emergence is optimal. One also expects the individual pupae of each species to exhibit a certain amount of variation about the mean so far as size, reserves, competency of the integuments and so on, is concerned. Some pupae will be slightly bigger, have slightly bigger reserves and more competent integuments. Yet others will be slightly smaller, have slightly smaller reserves and less competent integuments. To be succinct, one expects emergence to be Gaussian. 
\begin{assumption} \label{assumption4} 
{\bf \em The relationship between pupal emergence and humidity is a Gaussian curve, or a part thereof.} 
\end{assumption} 

At one extreme, one has environmentally highly specialised species with low hydrational inertia (e.g. {\em Glossina austeni}\footnotemark[1] and {\em G. brevipalpis}\footnotemark[1]) \footnotetext[1]{{\em G. austeni} has low hydrational inertia by virtue of its small size (which implies a high surface area to volume ratio), {\em G. brevipalpis} has low hydrational inertia due to inferior waterproofing.}, for which one expects variation over a small range in conditions to provide adequate data to fit the Gaussian curve. 

At the other extreme, species which exhibit massive hydrational inertia exist,
such as {\em G. longipennis} and {\em G. swynnertoni}. They provide little, or no clue as to
the underlying Gaussian relationship between emergence and humidity. The range 
of conditions, the domain, is not obviously suggestive of an underlying Gaussian
emergence curve. All one sees is a very small, consequently
flat-in-appearance, sample of the top of the curve. These species ought to have been investigated in terms of water loss rather than humidity. The
curves for other species lie between these extremes. 

In the wild, there is a compounding factor in that, not only is emergence based on variation within a given species, it is also based on variation within the environment. A whole range of microclimates exist within breeding sites, some of which are compost, rot holes and soil, to name only a few. Seasonal variation is a further compounding factor.

While the focus of this work is desiccation it is of interest to note that emergence also declines at very high humidities. As to whether drowning or some fungus is the desiderate explanation, it can only be speculated. 

A chi squared test is not expected to elucidate any more than visual inspection. Despite the impossibility of any rigorous hypothesis testing the author maintains that the assymptotic standard errors obtained in Table \ref{errors} make a compelling argument for Assumption \ref{assumption4}.
\begin{figure}[H]
    \begin{center}
\includegraphics[height=12cm, angle=-90, clip = true]{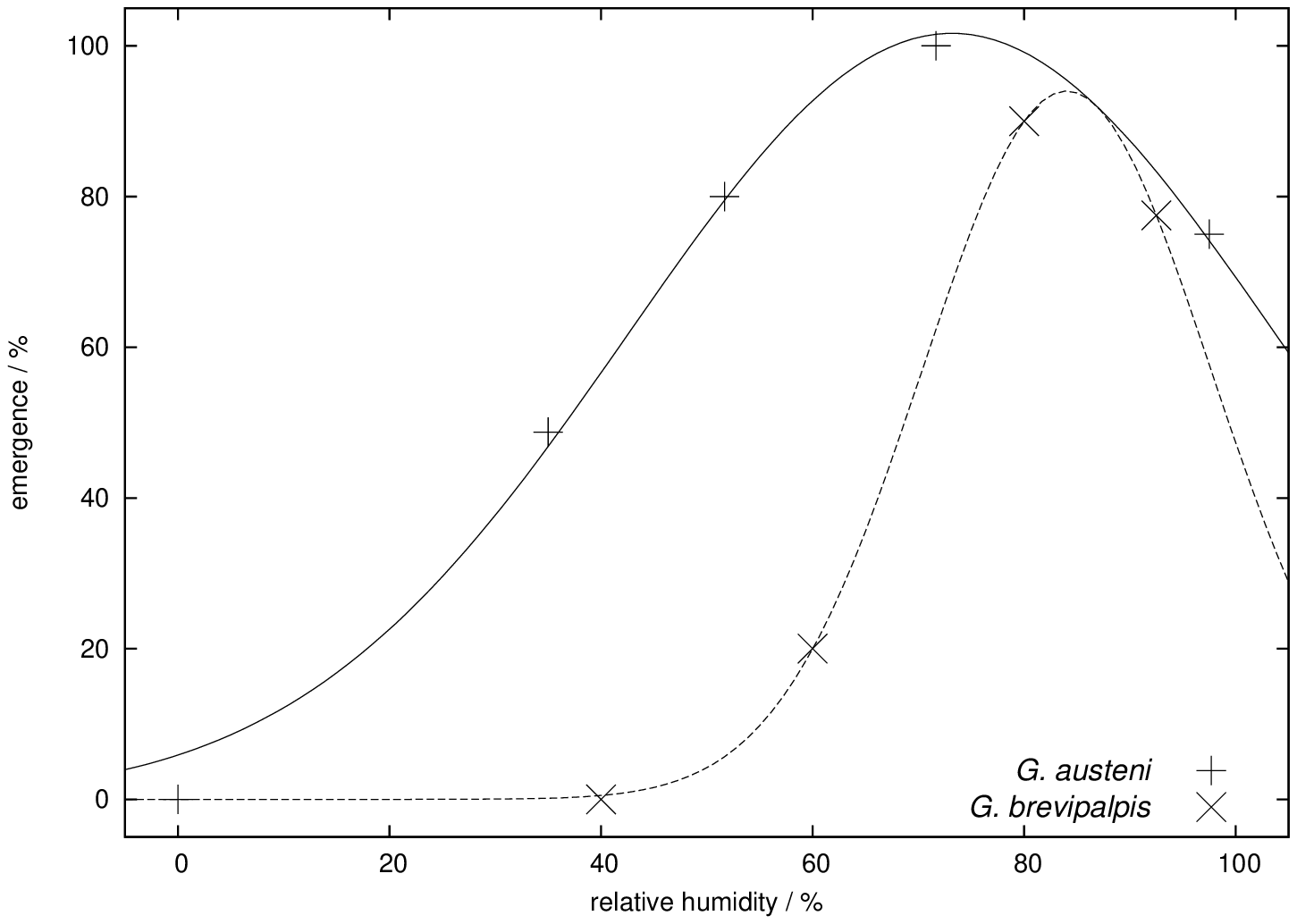}
\caption{Species with low hydrational inertia (e.g. {\em G. austeni} and {\em G. brevipalpis}) are good indicators of the underlying Gaussian relationship between emergence and humidity. (Particular attention is drawn to the data for the Brevipalpis fit in Table \ref{errors} on page \pageref{errors}.)} \label{lowHydrationalInertia}
   \end{center}
\end{figure} 
\begin{figure}[H]
    \begin{center}
\includegraphics[height=12cm, angle=-90, clip = true]{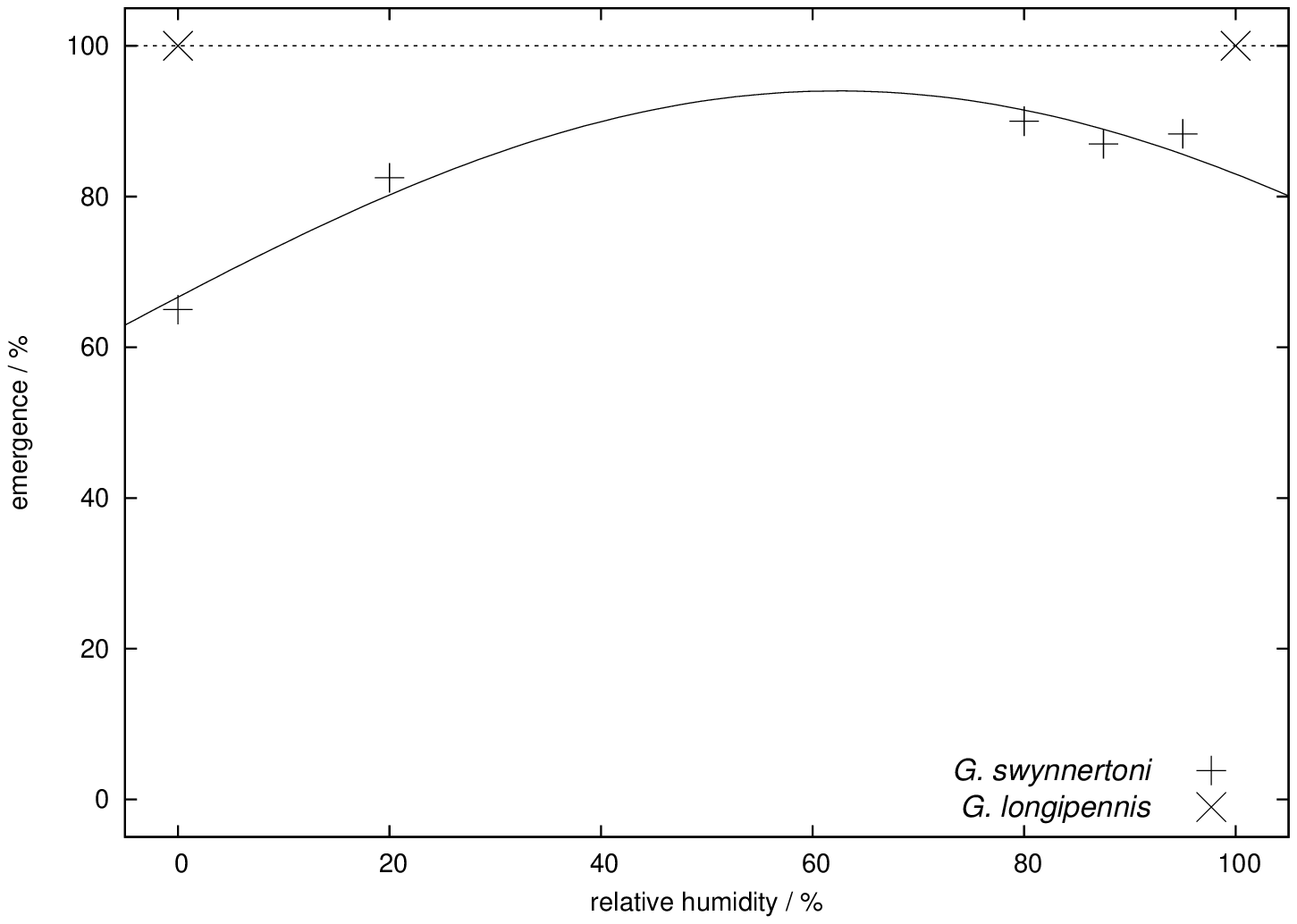}
\caption{Species with high hydrational inertia (e.g. {\em Glossina longipennis} and {\em Glossina swynnertoni}) provide little, or no clue as to the underlying Gaussian relationship between emergence and humidity.} \label{highHydrationalInertia}
   \end{center}
\end{figure}

\begin{table}
\begin{center}
\begin{tabular}{l l| c | c | c}  
&  & percentage & percentage assymptotic & sum of \\
group & species & emergence & standard error in & squares of \\ 
&  &  $E(h) = \displaystyle a \ e^{ - \frac{(h - b)^2}{2 c^2}}$  & $a$ \ \ \ \ \ \ \ \ \ \ \ \ $b$ \ \ \ \ \ \ \ \ \ \ \ \ $c$ & residuals \\ 
&  &  &  &  \\ \hline 
&  &  &  &  \\
{\em morsitans} & {\em austeni} \ & $101.663 e^{ - \frac{(h - 73.1591)^2}{2 \times 30.6468^2}}$ & 3.855 \ \ 2.077 \ \ 6.401 & 41.6495 \\ 
&  &  &  & \\ 
 \ & {\em morsitans} &  $94.4792 e^{ - \frac{(h - 70.6391)^2}{2 \times 77.3495^2}}$  &  5.177 \ \ 18.45 \ \ 29.07 & 305.227 \\ 
&  &  &  &  \\ 
 \ & {\em pallidipes} & $86.6257 e^{ - \frac{(h - 71.5636)^2}{2 \times 54.9713^2}}$ & 2.833 \ \ 6.205 \ \ 9.518 & 48.7588 \\ 
&  &  &  &  \\ 
 \ & {\em submorsitans} & $94.5092 e^{ - \frac{(h - 81.1895)^2}{2 \times 75.4474^2}}$ & 6.524 \ \ 37.45 \ \ 57.03 & 80.4127 \\ 
&  &  &  & \\ 
 \ & {\em swynnertoni} & $ 94.0194 e^{ - \frac{(h - 62.4064)^2}{2 \times 75.2339^2}}$ &  4.343 \ \ 8.65 \ \ 17.19 & 21.3715 \\ 
&  &  &  & \\ 
{\em palpalis} & {\em palpalis} & $95.8732 e^{ - \frac{(h - 78.8419)^2}{2 \times 23.4835^2}}$ & 13.89 \ \ 5.412 \ \ 23.94 & 725.514 \\
&  &  &  & \\ 
 \ & {\em tachinoides} & $98.8383 e^{ - \frac{(h - 79.6877)^2}{2 \times 40.8616^2}}$ & 11.02 \ \ 17.66 \ \ 29.46 & 427.406 \\ 
&  &  &  & \\ 
{\em fusca} & {\em brevipalpis} \ &  $94.0057 e^{ - \frac{(h - 84.0199)^2}{2 \times 13.6433^2}}$  & 0.5123 \ \ 0.1352 \ \ 0.907 & 0.268286 \\ 
&  &  &  & \\ 
\end{tabular}
\caption{The percentage emergence, the assymptotic standard errors and the sum of squares of the residuals for the fit in each species.} \label{errors}
\end{center}
\end{table}

\begin{figure}[H]
    \begin{center}
\includegraphics[height=16cm, angle=-90, clip = true]{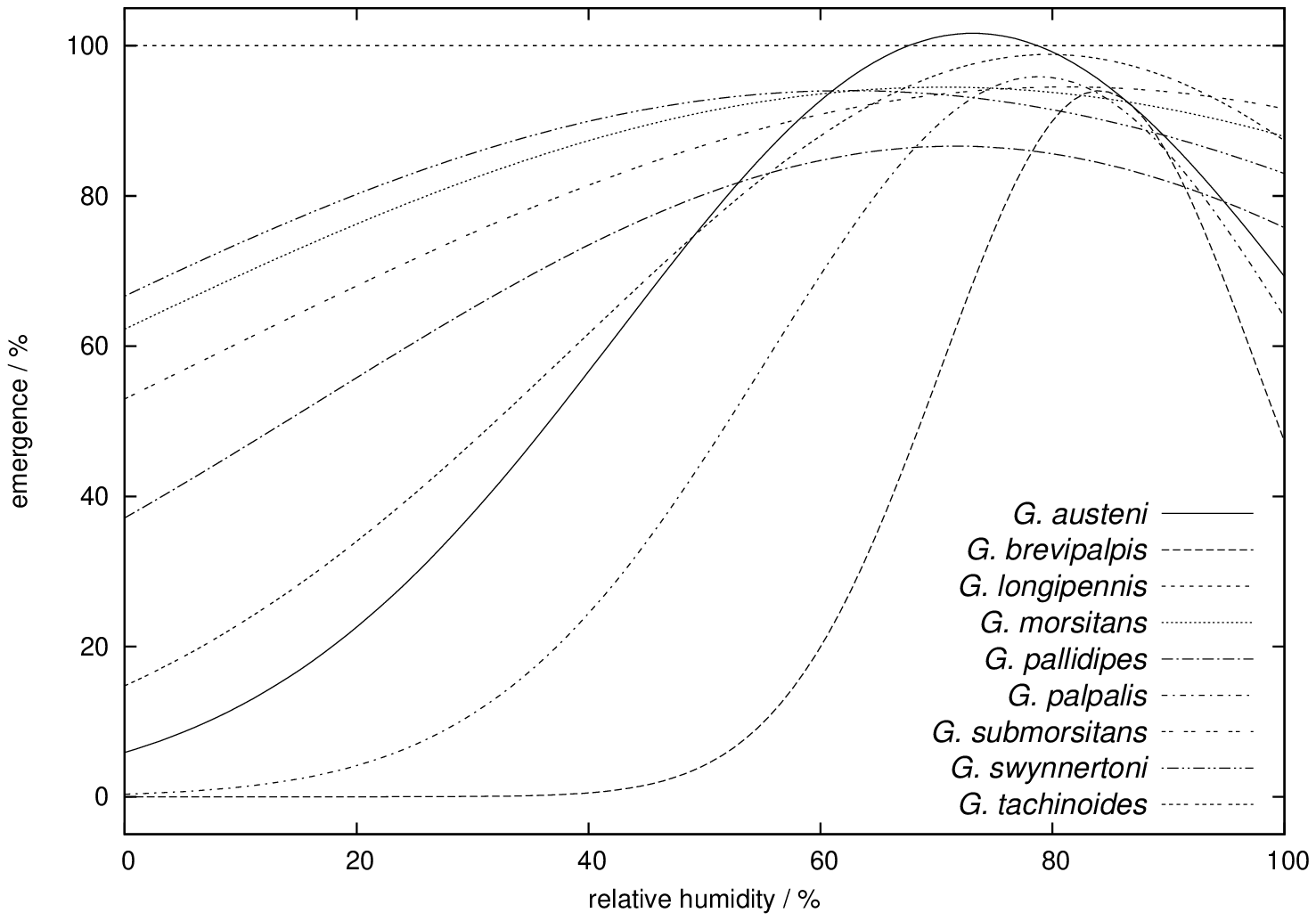}
\caption{Percentage emergence modelled as a Gaussian curve$^1$ for a variety of species. All are at 24$^o$C, except {\em G. tachinoides} (30$^o$C).} \label{allSpeciesTogether}
   \end{center}
\end{figure}
\footnotetext[1]{{\em G. longipennis} is the single exception (a straight line had to be fitted to the only two data points).}

\subsubsection*{The Issue of Sub-Standard, Laboratory Pupae}

In {\sc Bursell} \cite{Bursell1} it is somewhat heuristically argued that the laboratory pupae in question were too small and that all emergence curves should therefore be displaced 10\% to the left (the right in those graphs). An alternative argument based on puparial transpiration and in which the pupa is approximated as a spheroid, is preferred (on page \pageref{subStandardPupae} of the addendum). It entertains replacing the emergence function, $E(h)$, with 
\begin{eqnarray} \label{21}						
E \left( \frac{h + 3.57}{\sqrt[3]{\frac{100}{90}}} \right)
\end{eqnarray}		
as an alternative. 

\subsection{Survival to Emergence for Historically Variable Humidities and Temperatures}

Problems, similar to those encountered for historical water loss, compound matters when it comes to the survival to eclosion for each species. Historical humidities are steady, furthermore, the data were obtained at a constant 24$^o$C. 
\begin{assumption} \label{assumption5} 
{\bf \em The pupal emergence for a given water loss, is the same as the pupal emergence for a steady humidity at 24$^{\mbox{o}}$C, that produced an equivalent total water loss.} 
\end{assumption} 
In other words, it is assumed that emergence can be re-expressed in terms of total water loss. Do different histories in temperature and humidity, which produce the same water loss, imply the same pupal emergence, or is the amount of water present at some particular stage more relevant to the pupa's survival to full term? The simple answer is to refer the reader to the title, although this does somewhat avoid the question.

In practice, it is far easier to convert a total computed water loss to a corresponding steady-humidity-at-24$^o$C\footnotemark[1], instead of the other way around \footnotetext[1]{G. tachinoides data the exception, being at 30$^o$C}. Either way entertains the possibility of fictitious, or negative, humidities. (There are always those who are apt to find this sort of thing vaguely disturbing, however, it should be pointed out that $E(h)$ is just a mathematical function and $h$, a variable. One really needs to think `outside the box' in these matters. What was humidity is now not so much an artefact, just something a little more abstract.)

The results of the water loss algorithm for 24$^o$C and any, given set of steady humidities obviously constitute a monotonic decline. The problem, however, is that the complete algorithm is relatively involved and voluminous. Under these circumstances, practical considerations and not rates of convergence dictate implementation of an 1/2 interval search. (The rate of convergence is not bad in this instance.)

\section{Remaining Fat Reserves}

Remaining fat reserves are a more tenuous result. Water content stays constant after the 3rd and 4th instars {\sc Glasgow} \cite{Glasgow1}. Although the oxidation of fat for the specific purpose of water production is suspected, it could not be proven at a statistically significant level ({\sc Bursell} \cite{Bursell1}). The conversion factor (by mass) is given as 
\begin{eqnarray*}
\mbox{fat oxidised} &=& 1.12 \times \mbox{water}. 
\end{eqnarray*} 

\section{Testing the Model}

Testing the model presents something of a challenge. The emergence data at
24$^o$C, to a certain extent, provides a test for self-consistency. Transecting
the following surfaces of emergence at 24$^o$C should come close to replicating
the Gaussian curves for emergence, on page \pageref{allSpeciesTogether}.
``Close'', since the data were adjusted as a consequence of the inferior pupae
issue.

Testing for corroboration with observed mortalities in the field is somewhat heuristic. All one can say is that predicted pupal mortalities due to water loss should, logically, never exceed any pupal mortalities observed in the field for similar conditions of humidity and temperature. For example, the computed {\em G. morsitans} emergence due to water loss is not lower than the 87\% obtained by {\sc Hargrove and Williams} \cite{HargroveAndWilliams} in their Antelope Island mark-recapture experiment.

One set of data on which the model was not based does, however, still remain. Those data are the measured initial water reserve for a number of species. It has the makings of a test for consistency with reserves. In a perfect world, the measured, critical water loss contour should correspond to that of this model's median, or 50\%, emergence contour. {\em G. morsitans} is of obvious interest as the species on which the model is based. 
\begin{figure}[H]
    \begin{center}
\includegraphics[height=11cm, angle=0, clip = true]{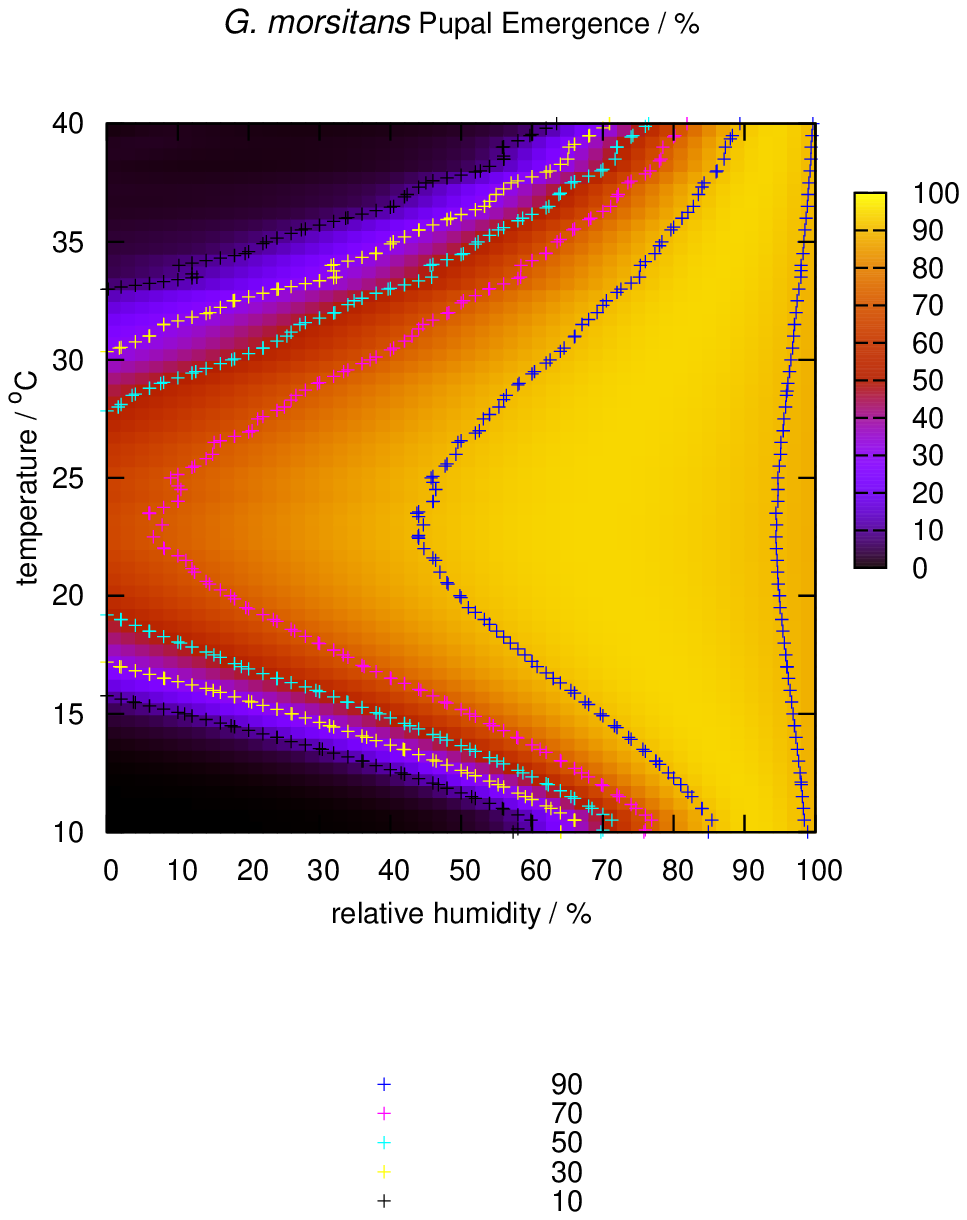}
\includegraphics[height=11cm, angle=0, clip = true]{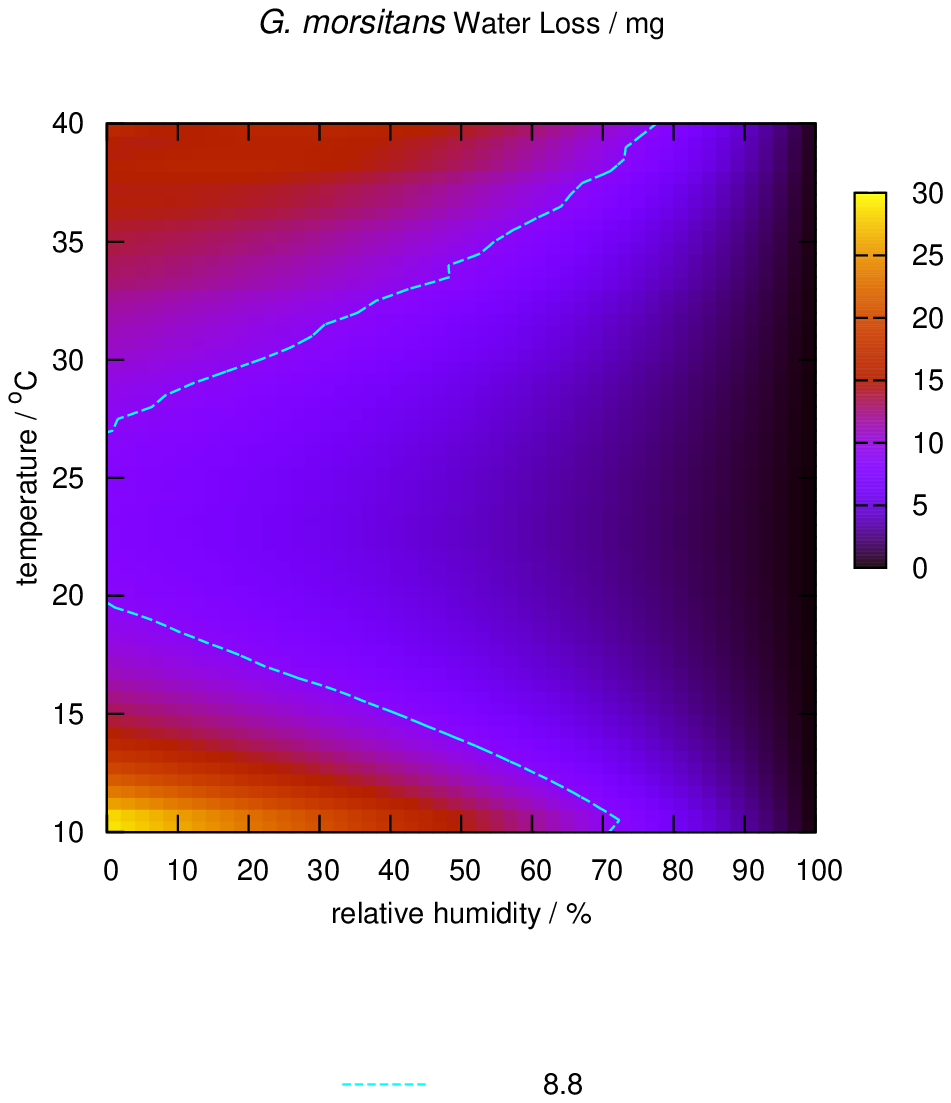}
\caption{Computed pupal emergence (left) and water loss (right) for {\em G. morsitans}.} \label{morsitansEmergenceAndWaterLoss}
   \end{center}
\end{figure} 

\begin{figure}[H]
    \begin{center}
\includegraphics[height=11cm, angle=0, clip = true]{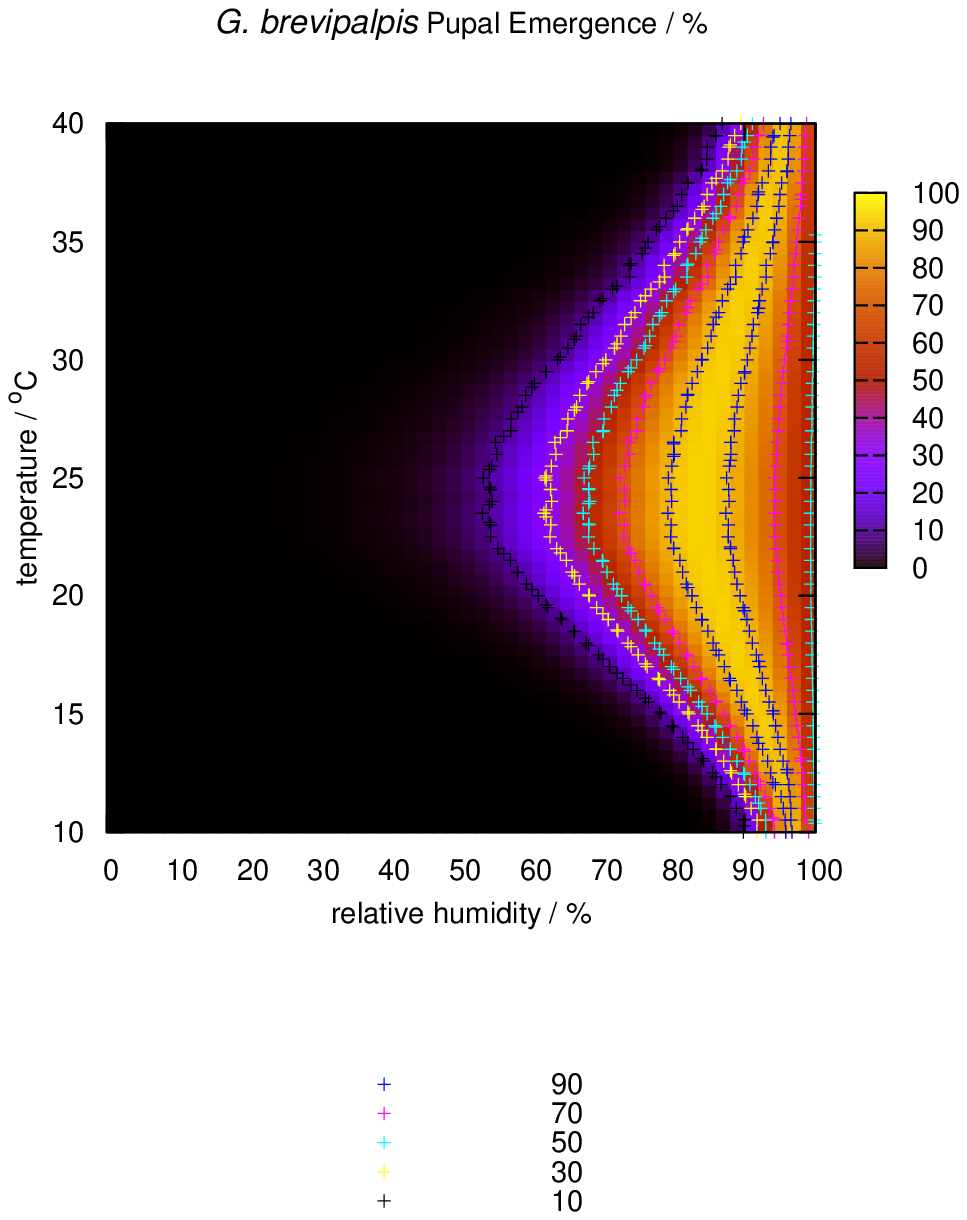}
\includegraphics[height=11cm, angle=0, clip = true]{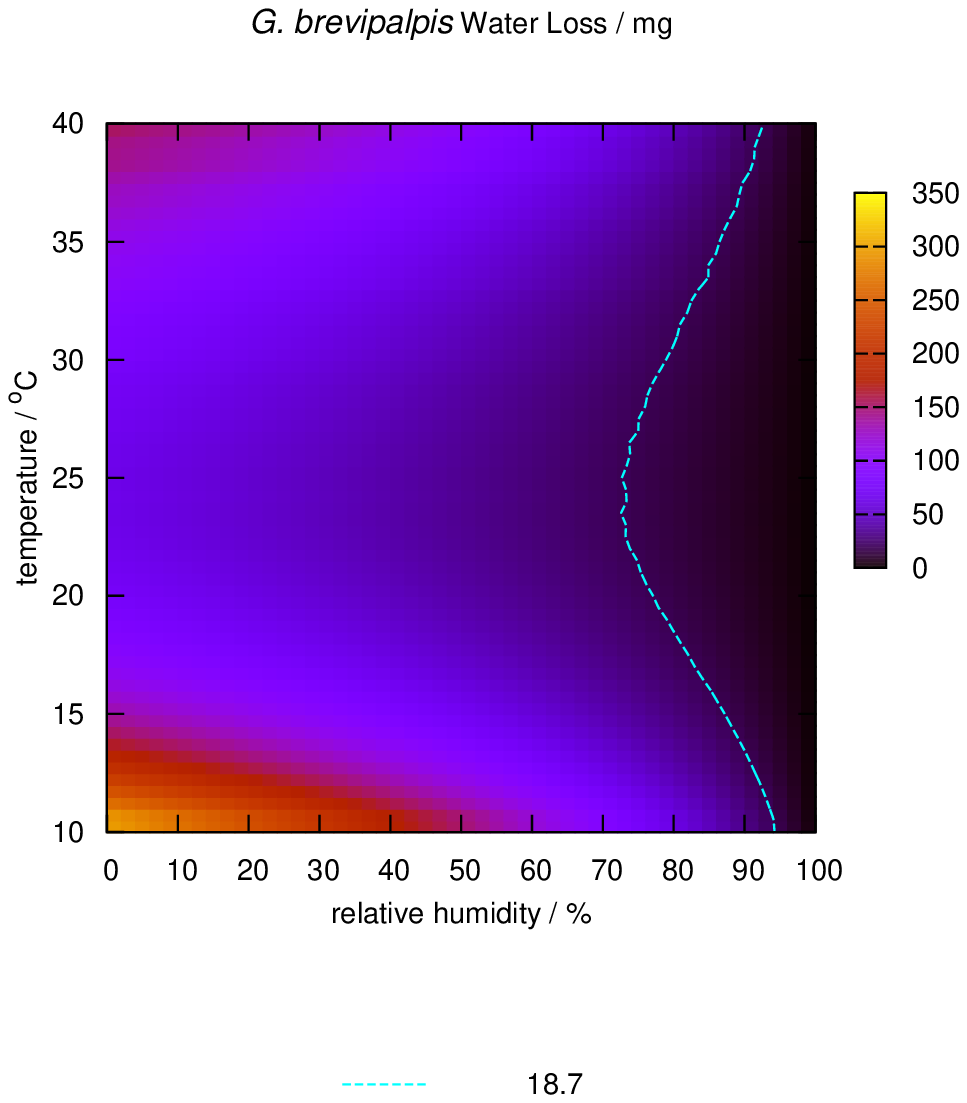}
\includegraphics[height=11cm, angle=0, clip = true]{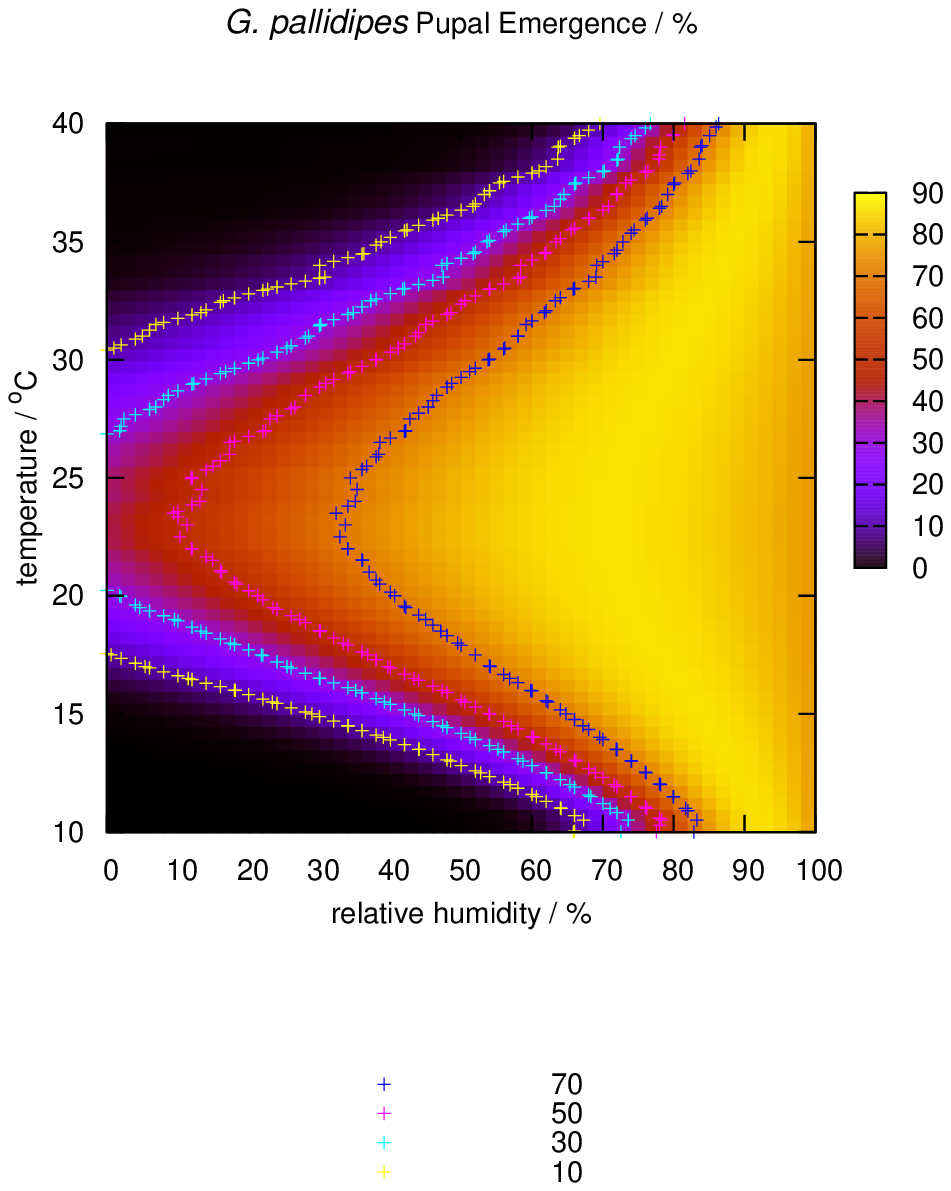}
\includegraphics[height=11cm, angle=0, clip = true]{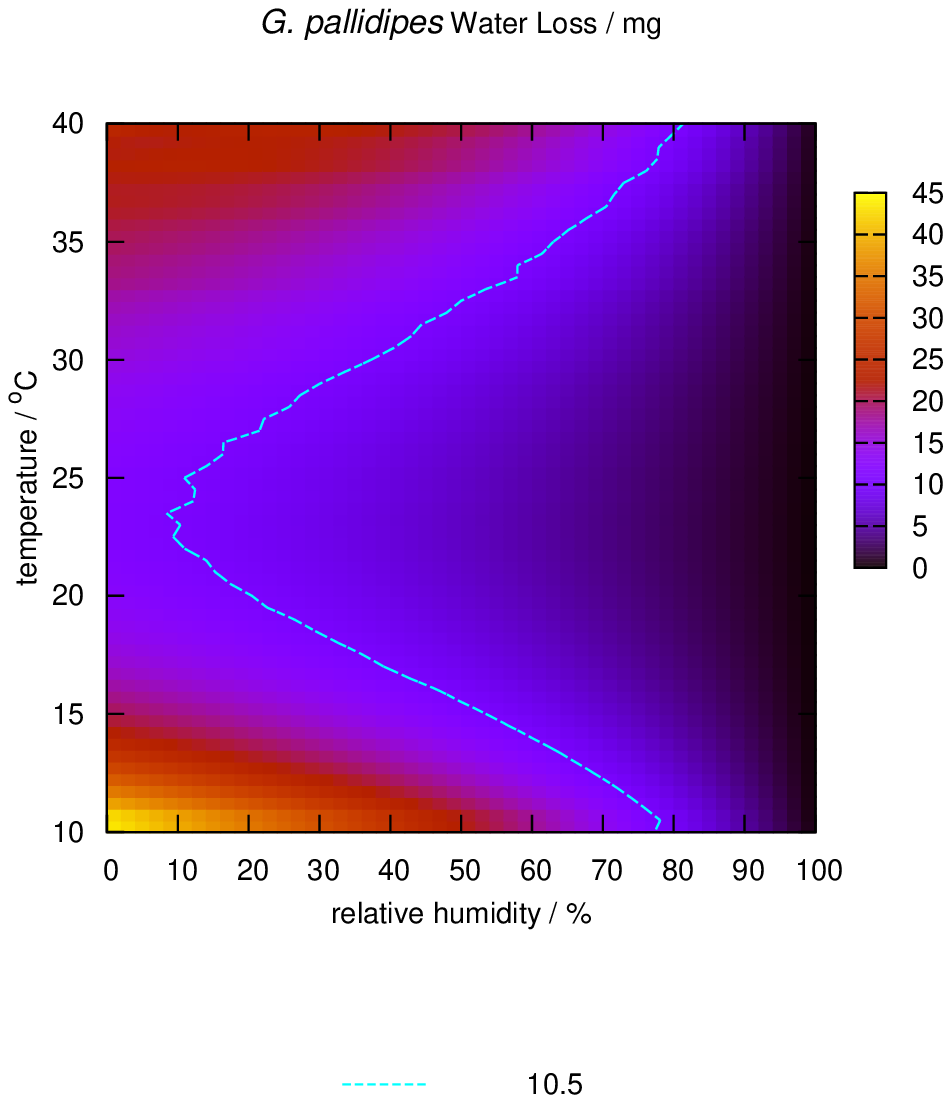}
\caption{Computed pupal emergence (top left) and water loss (top right) for {\em G. brevipalpis}; computed pupal emergence (bottom left) and water loss (bottom right) for {\em Glossina pallidipes}.} \label{brevipalpisAndPallidipes}
   \end{center}
\end{figure} 
\begin{figure}[H]
    \begin{center}
\includegraphics[height=11cm, angle=0, clip = true]{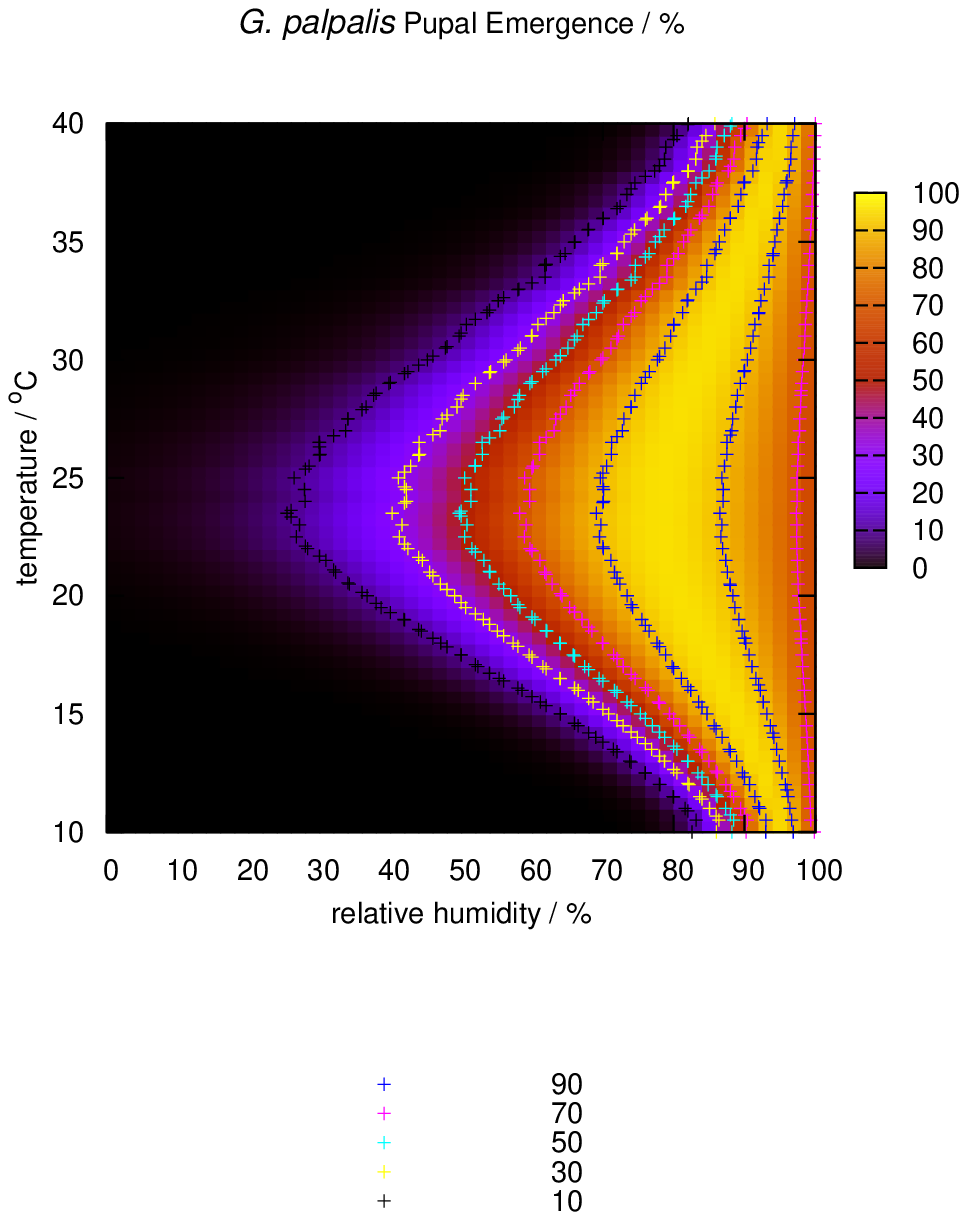}
\includegraphics[height=11cm, angle=0, clip = true]{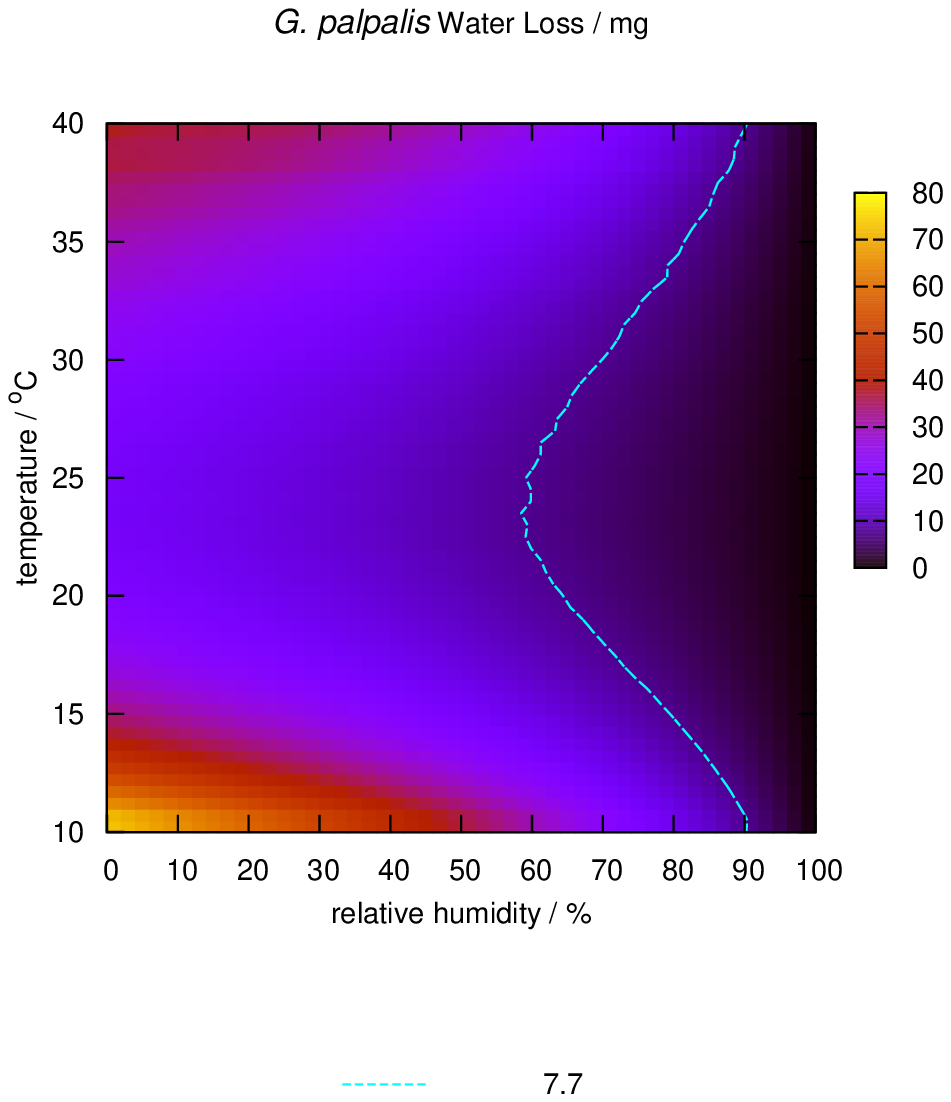}
\includegraphics[height=11cm, angle=0, clip = true]{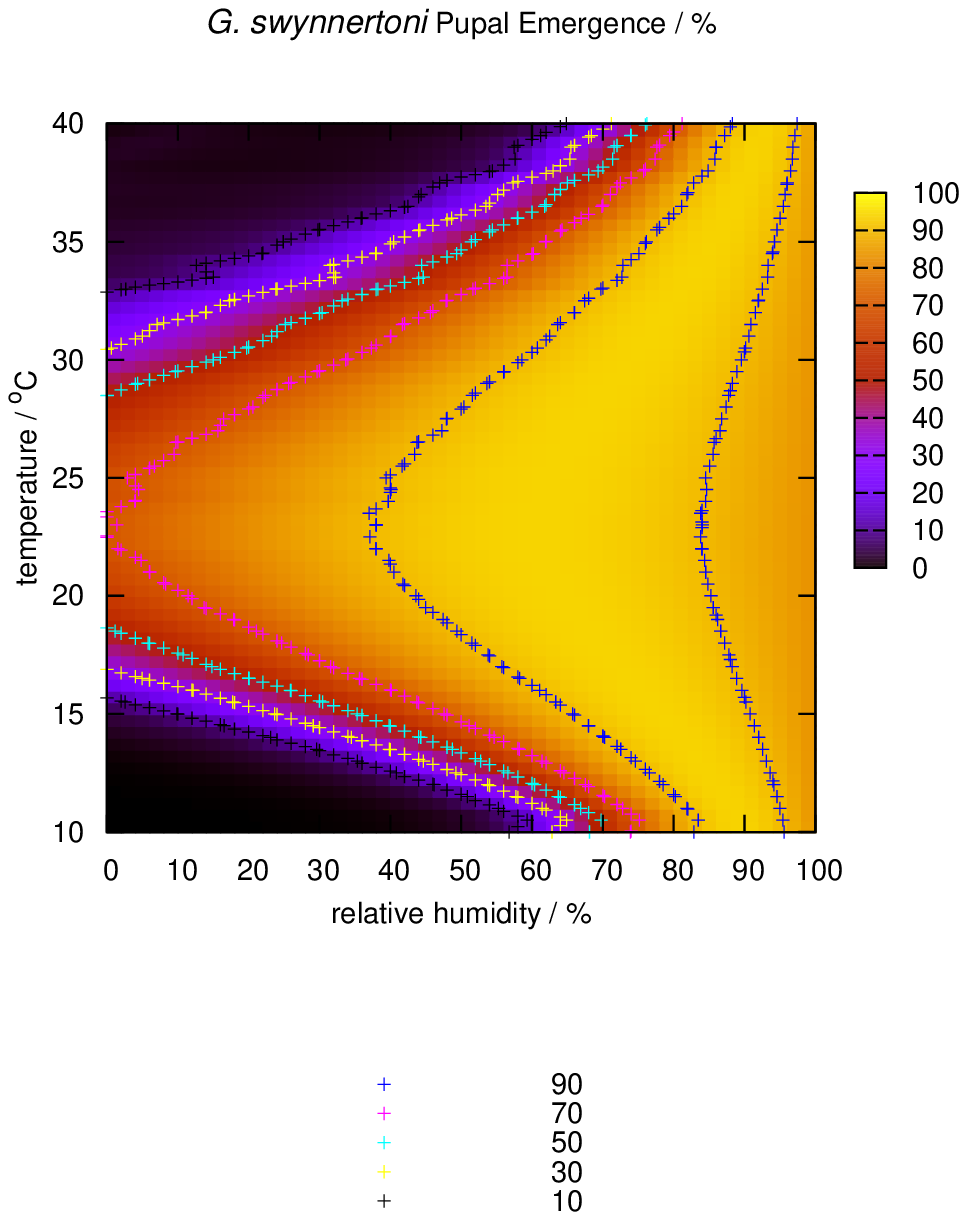}
\includegraphics[height=11cm, angle=0, clip = true]{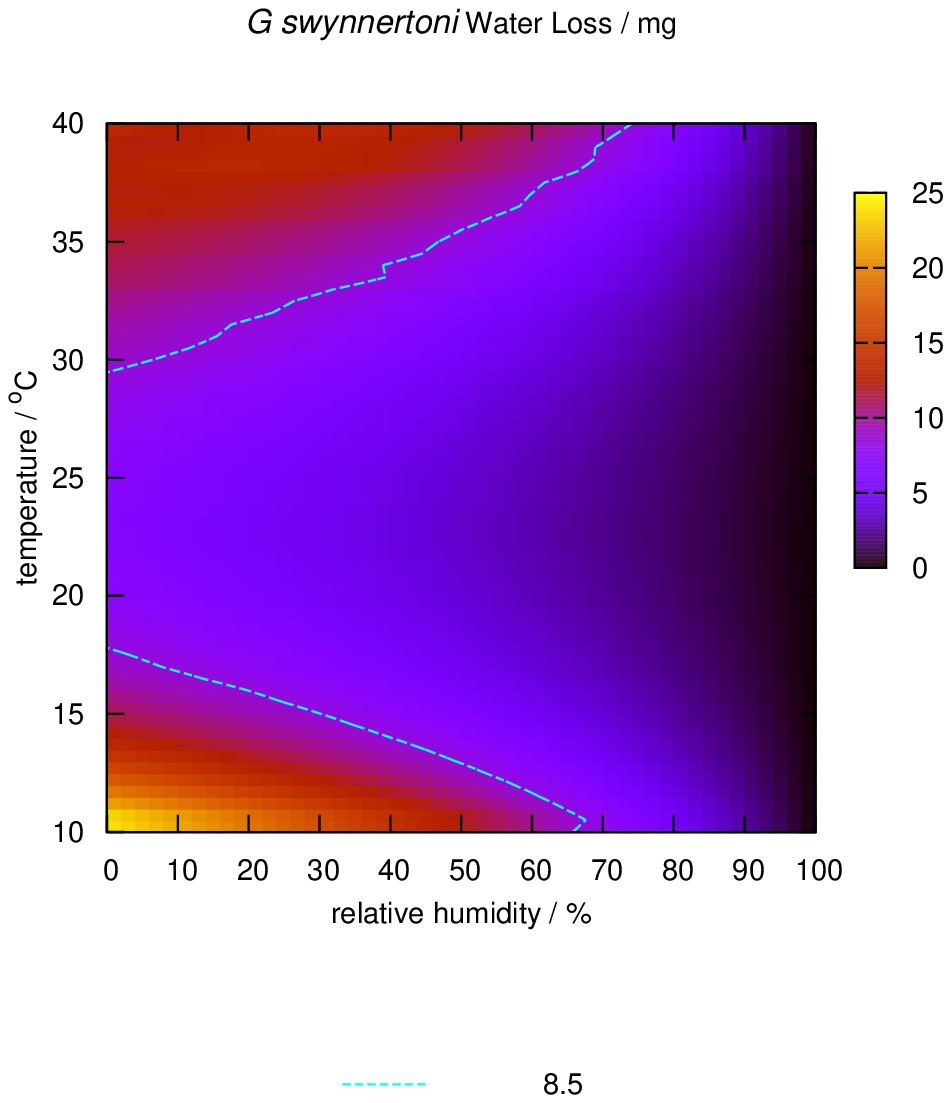}
\caption{Computed pupal emergence (top left) and water loss (top right) for {\em G. palpalis}; computed pupal emergence (bottom left) and water loss (bottom right) for {\em G. swynnertoni}.} \label{palpalisAndSwynnertoni}
   \end{center}
\end{figure} 
The sensitivity of {\em G. brevipalpis} and {\em G. palpalis} pupae to dehydration makes these species arguably the most challenging tests, as well as of particular interest to this work. {\em G. brevipalpis} is, furthermore, of national interest to those who funded this work and {\em G. palpalis} is known to feed on humans in West Africa ({\sc Solano} \cite{Solano1}), it being a major culprit in the spread of human trypanosomiasis. In both cases, information on the respective 4th instar excretions is available. 

The critical water reserve is also known for two other species which are expected to be less challenging as tests and for which {\em G. morsitans}-type, 4th instar excretions are expected to suffice. The water loss associated with the 4th instar excretions only becomes relevant close to the dew point (a part of the domain in which we have little interest) and it is, furthermore, thought to be too small to be of any real consequence.
\begin{table}[H]
\begin{center}
\begin{tabular}{c c c c c}  
{\em G. brevipalpis} \ \ & \ \ {\em G. morsitans} \ \ & \ \ {\em G. pallidipes} & \ \ {\em G. palpalis} & \ \ {\em G. swynnertoni} \\
&  &  &  &  \\ 
18.7mg & 8.8mg & 10.5mg & 7.7mg & 8.5mg \\ 
\end{tabular}
\end{center}
\caption{Initial water reserves after {\sc Bursell} \cite{Bursell1}.} \label{reserves}
\end{table}

Given these initial water reserves, the position of the critical water loss
contour should correspond to that of the median, or 50\%, emergence contour for
the model to work (Figures \ref{morsitansEmergenceAndWaterLoss} to
\ref{palpalisAndSwynnertoni}). In this regard, it is worth noting that the {\em
G. morsitans}-based puparial duration used in the model could be as little as
83\% of the {\em G. brevipalpis} puparial duration and the correspondence
observed in the Figure \ref{brevipalpisAndPallidipes} result should therefore
not be as good as it is. 

\section{Conclusions}

This work gives rise to a sequence of governing equations, an algorithm and an
applet. They predict water loss, consequent daily mortality and percentage
emergence due to water loss, given the species and the variable daily
temperatures and humidities which prevailed during pupation. In
this way it is hoped that the model brings a certain degree of closure to the
question of pupal dehydration in tsetse. The results are certainly adequate to
conclude proof-of-concept for the model, at very least. The model also
provides a certain amount of insight.

High transpiration rates are a consequence of high temperatures and unfavourable
humidities. They lead to a dehydration of the tsetse pupa which can be fatal.
Although the diametrical opposite is true of transpiration rates at low
temperatures, metabolic processes are slowed, the puparial duration becomes too
long and the cumulative effect of transpiration can be just as fatal. 

It would appear that the similarities between the various {\em Glossina} species
are better than expected so far as pupal water loss is concerned. The
correspondence between actual measured, critical water losses and water losses
calculated based on Assumption \ref{assumption3}, is profound and no discernable
error exists for some species e.g. {\em G. pallidipes} and {\em G. swynnertoni}.
\begin{figure}[H]
    \begin{center}
\includegraphics[height=11cm, angle=0, clip = true]{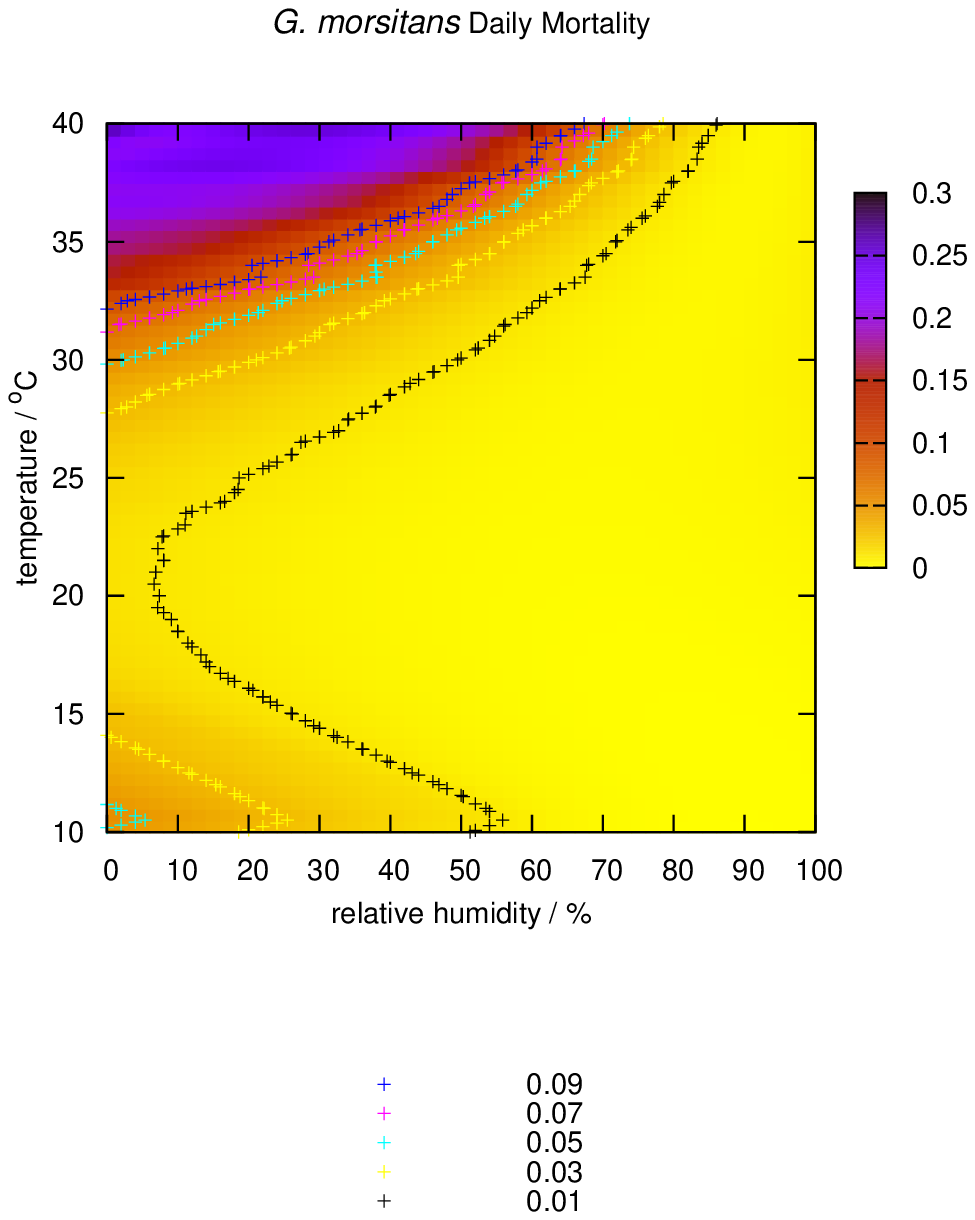}
\includegraphics[height=11cm, angle=0, clip = true]{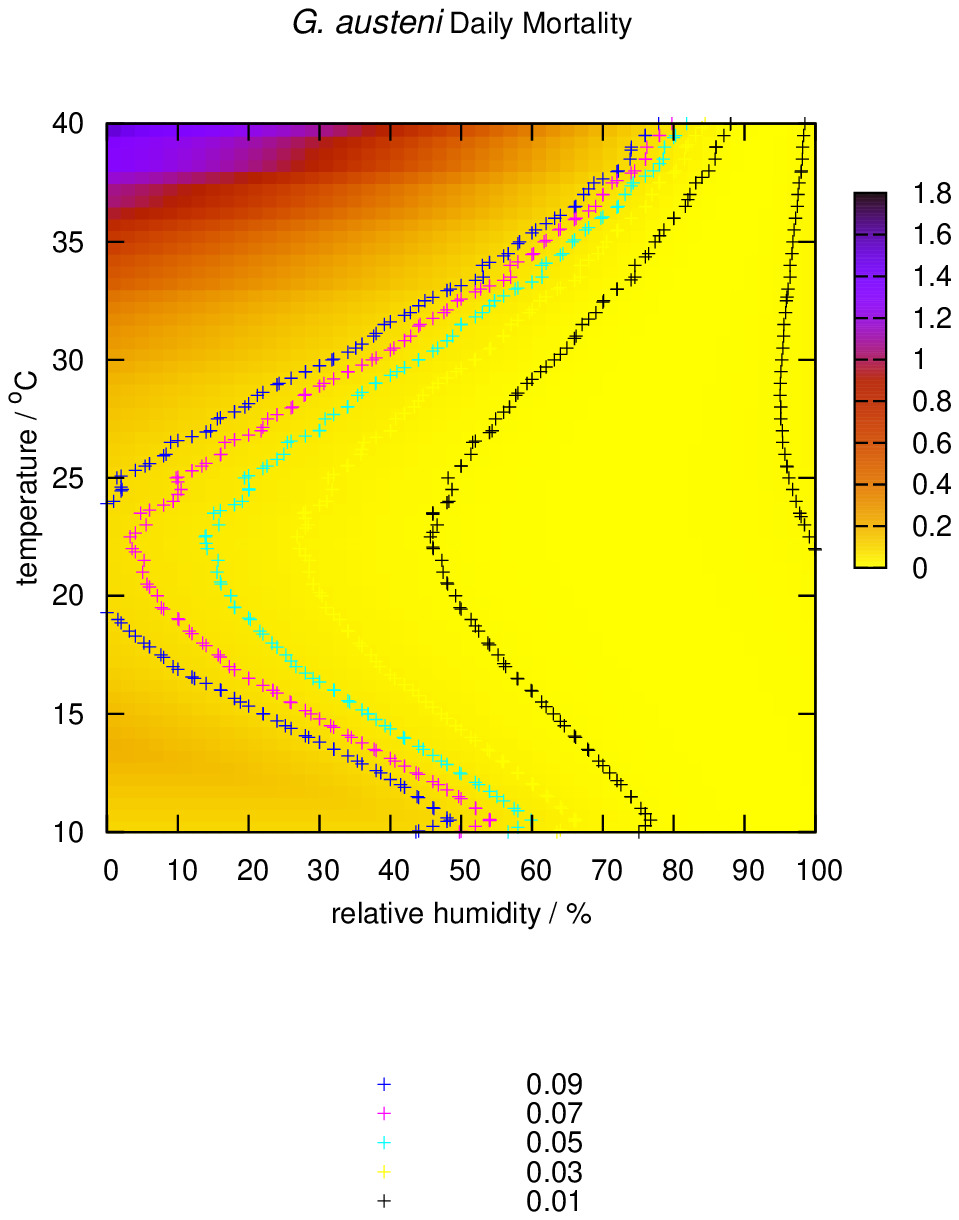}
\includegraphics[height=11cm, angle=0, clip = true]{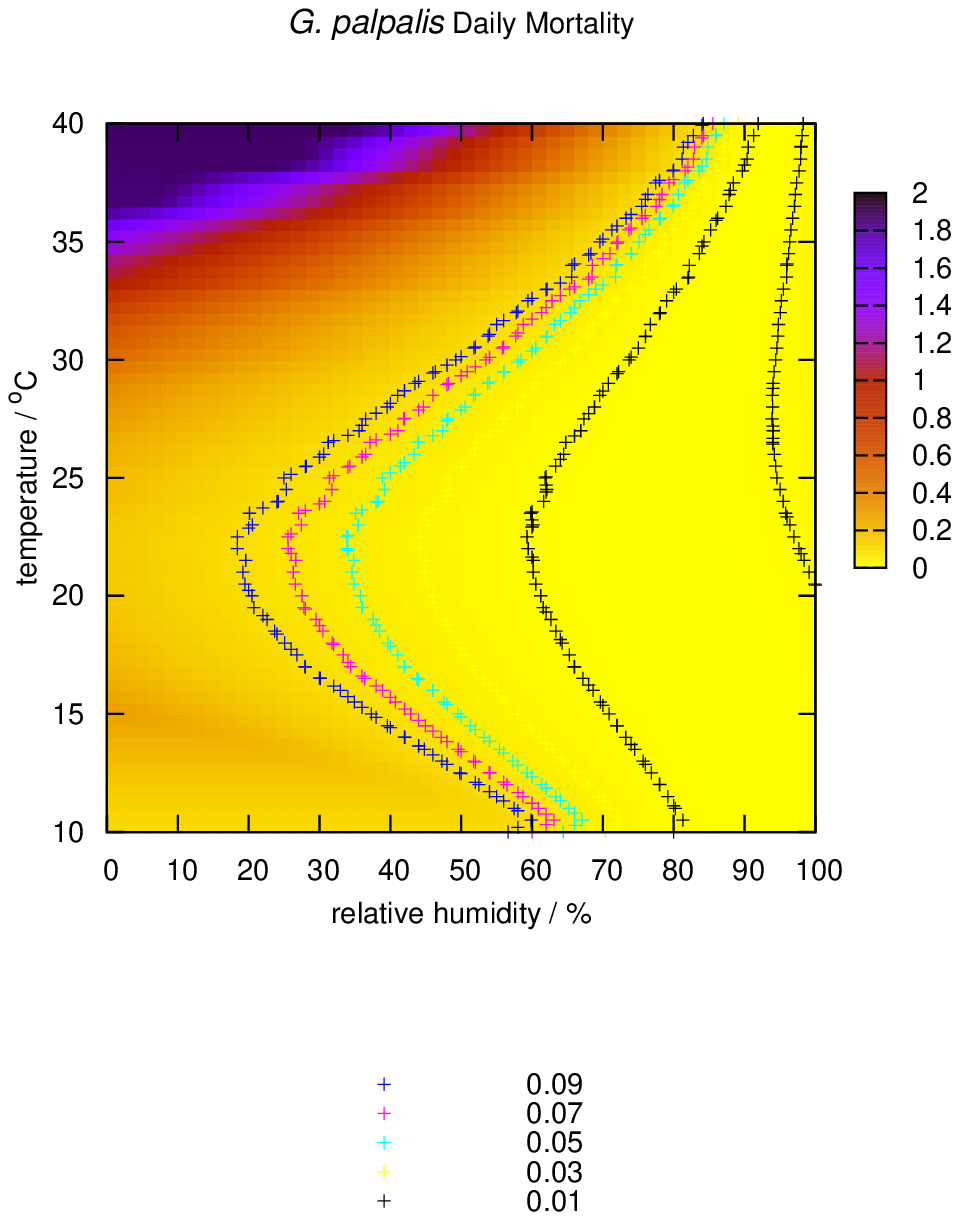}
\includegraphics[height=11cm, angle=0, clip = true]{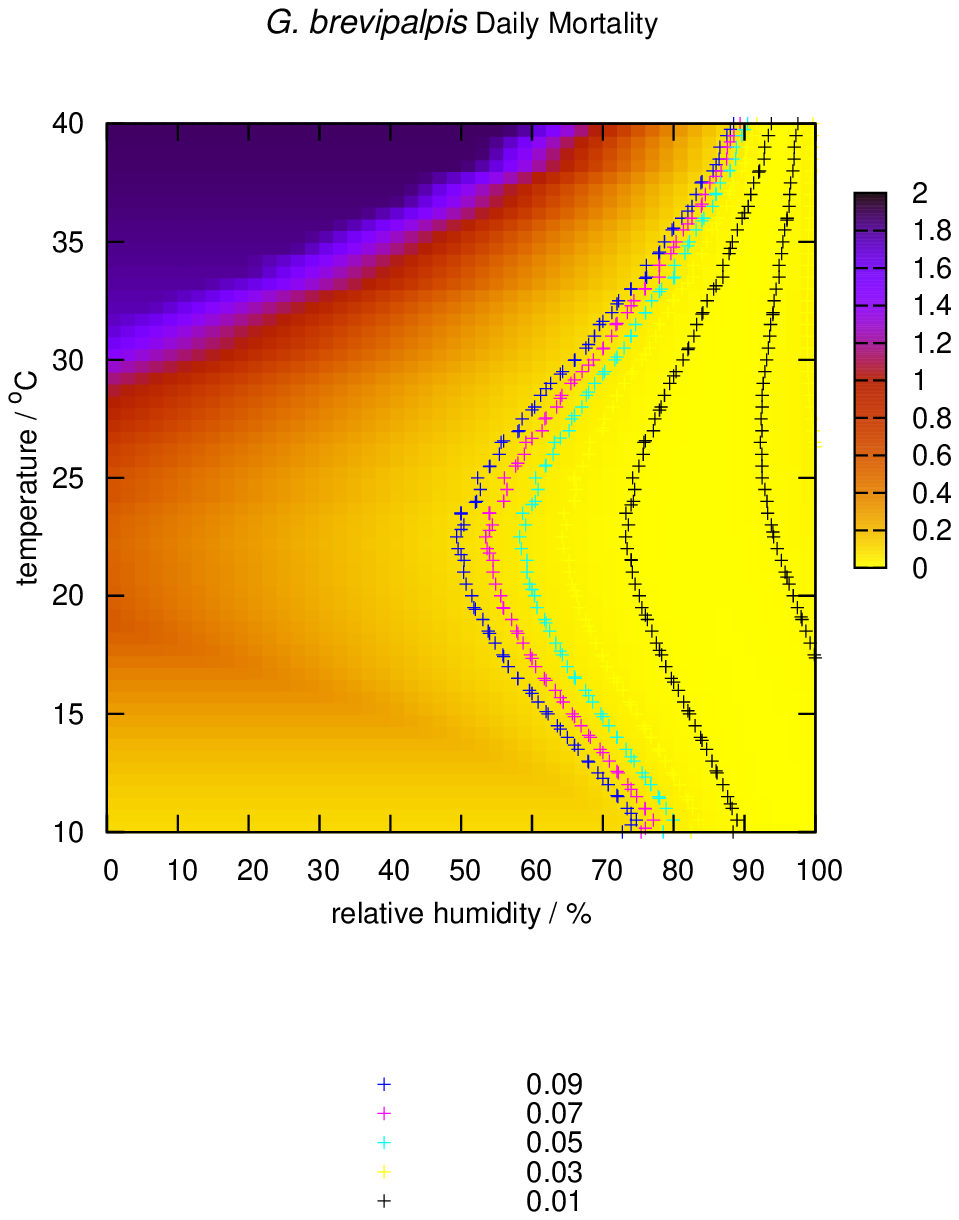}
\caption{Computed daily mortality for {\em G. morsitans} (top left), {\em G. austeni} (top right), {\em G. palpalis} (bottom left), {\em G. brevipalpis} (bottom right).} \label{mortality}
   \end{center}
\end{figure} 
It would therefore appear that different species of pupae behave fairly
similarly in the ground, that they actively pursue a strategy to minimise water
loss for the majority of modern habitats and have hydrational mechanisms
preventative of desiccation. The {\em Glossina} genus therefore may well derive
from a common, tropical, rain-forest dwelling ancestor, adjusted to moist, warm
climates and dehydration is a challenge to pupae. The fact that loss rates in
{\sc Bursell} \cite{Bursell1} were originally determined as fractions of initial
pupal masses, rather than per unit of surface area, therefore turned out to be a
windfall rather than a criticism (defended in {\sc Bursell} \cite{Bursell2}).
The necessary information to convert between the two turns out to facillitate
the adaption of the {\em G. morsitans} model to other species. 

A strong school of thought is of the opinion that pupal and teneral mortality
due to dehydration are either irrelevant or can be assumed constant. Do daily,
pupal mortalities due to water loss justify this work? Figure \ref{mortality} is
suggestive of what the origins of such an argument might be in the case of the
{\em morsitans} group, even for {\em G. austeni}\footnotemark[1]
\footnotetext[1]{{\em G. austeni} is thought to have secondarily invaded moist
forest areas (it is small, with all the lack of hydrational inertia that a high
surface area to volume ratio implies).}. Notice, however, that even for a hardy
fly such as {\em G. morsitans}, its prospects deteriorate rapidly once out of
favourable habitat. Even for the {\em morsitans} group, daily pupal mortality is
neither linear, nor a function of temperature alone. When it comes to some
members of the {\em fusca} and {\em palpalis} groups, however, that pupal
mortality due to dehydration is both relevant and palpable is beyond contention.
The effect of humidity is profound. Humidity defines habitat. This is despite
the fact that water loss and any consequent pupal mortality are also very
different things (one expects water loss to culminate in, and ultimately take
its toll on the teneral). In this regard, it is of interest to note that {\sc
Rogers} and {\sc Robinson} \cite{RogersAndRobinson} found that cold cloud
duration (rainfall) was far and away the most frequently occurring variable in
their top five for determining the distribution of both the {\em fusca} and {\em
palpalis} groups using satellite imagery. Normalized difference vegetation index
(NDVI) ranked second, again, by a significant margin. It is not too great
a stretch of the imagination to entertain the possibility that cold cloud
duration and NDVI translate directly into soil humidity, as might
elevation in the context of low-lying, coastal areas. (They also found that
rainfall was even more relevant when it came to abundance, as opposed to
distribution.)

Two gender selection effects are evident based on puparial duration alone. Under
adverse conditions more females than males emerge. A less significant,
male-selective phenomenon occurs close to dewpoint. Both phenomena are expected
to be enhanced should the heavier female mass and lower relative surface area
be taken into account (no data could be found).

The possibility that plots presented in this work may point to the fact that breeding sites for some species could be predicted to be very much confined in the dry season, cannot be ruled out. These would be obvious places in which to concentrate traps and one immediate application of this work.

It was stated in the introduction that early stage mortality is considered to be
the most significant, by far, in any model of tsetse population dynamics and
that a cursory inspection of the literature suggests pupal dehydration to be the
most challenging aspect of modelling it. The prognosis for this rather
simplistic, experimental model would therefore be one of greater significance
and any problems arising from issues such as inferior quality pupae, differing
puparial durations and the shortage of statistically significant data can be
corrected at some stage. One would like to believe in a more meaningful
accomplishment: That the completion of this work could be supposed to leave the
way open to a comprehensive model of early mortality, based either on a joint
probability density function, or, more likely, a Markov chain. 
\begin{figure}[H]
   \begin{center}
\includegraphics[height=11cm, angle=0, clip = true]{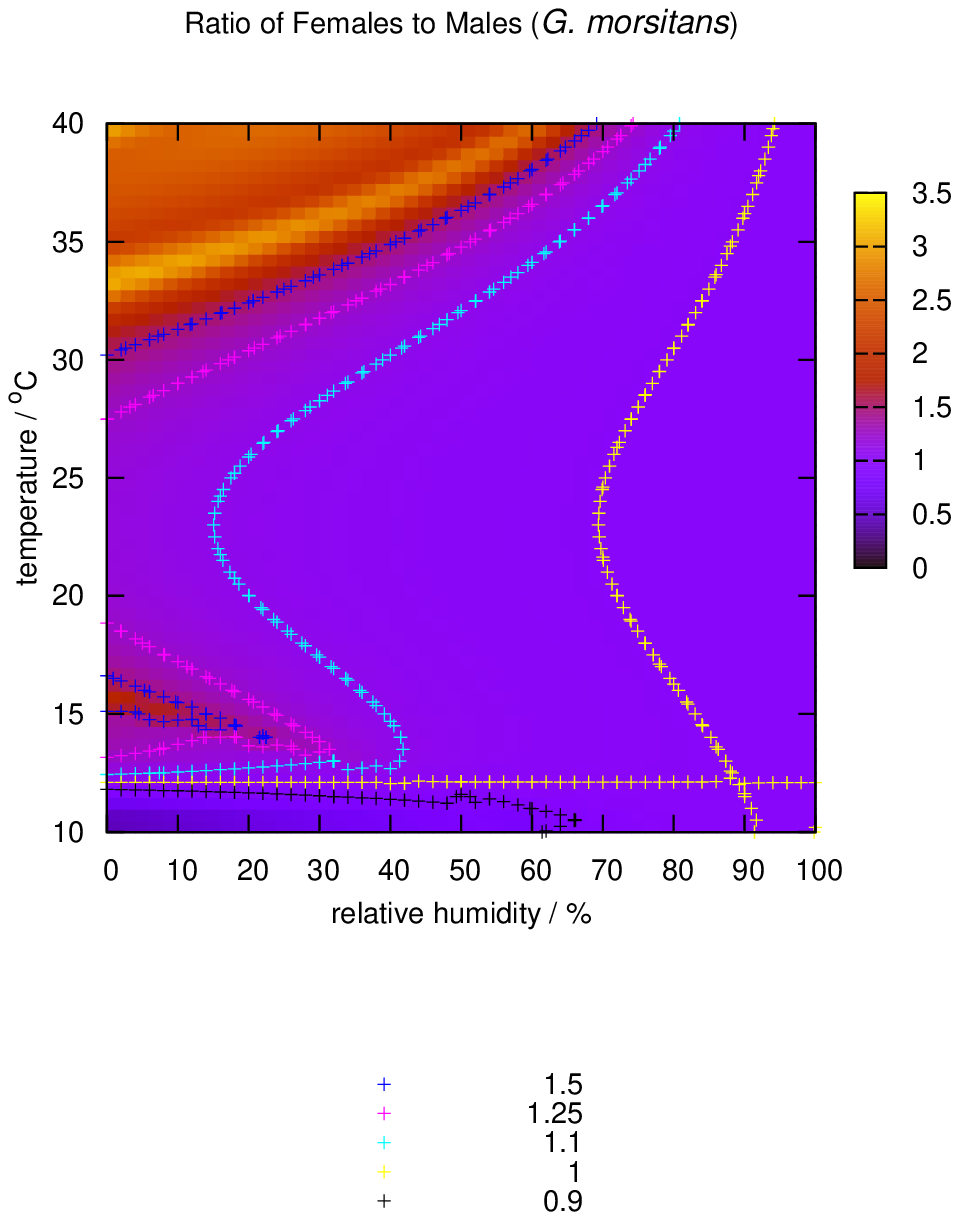}
\includegraphics[height=11cm, angle=0, clip = true]{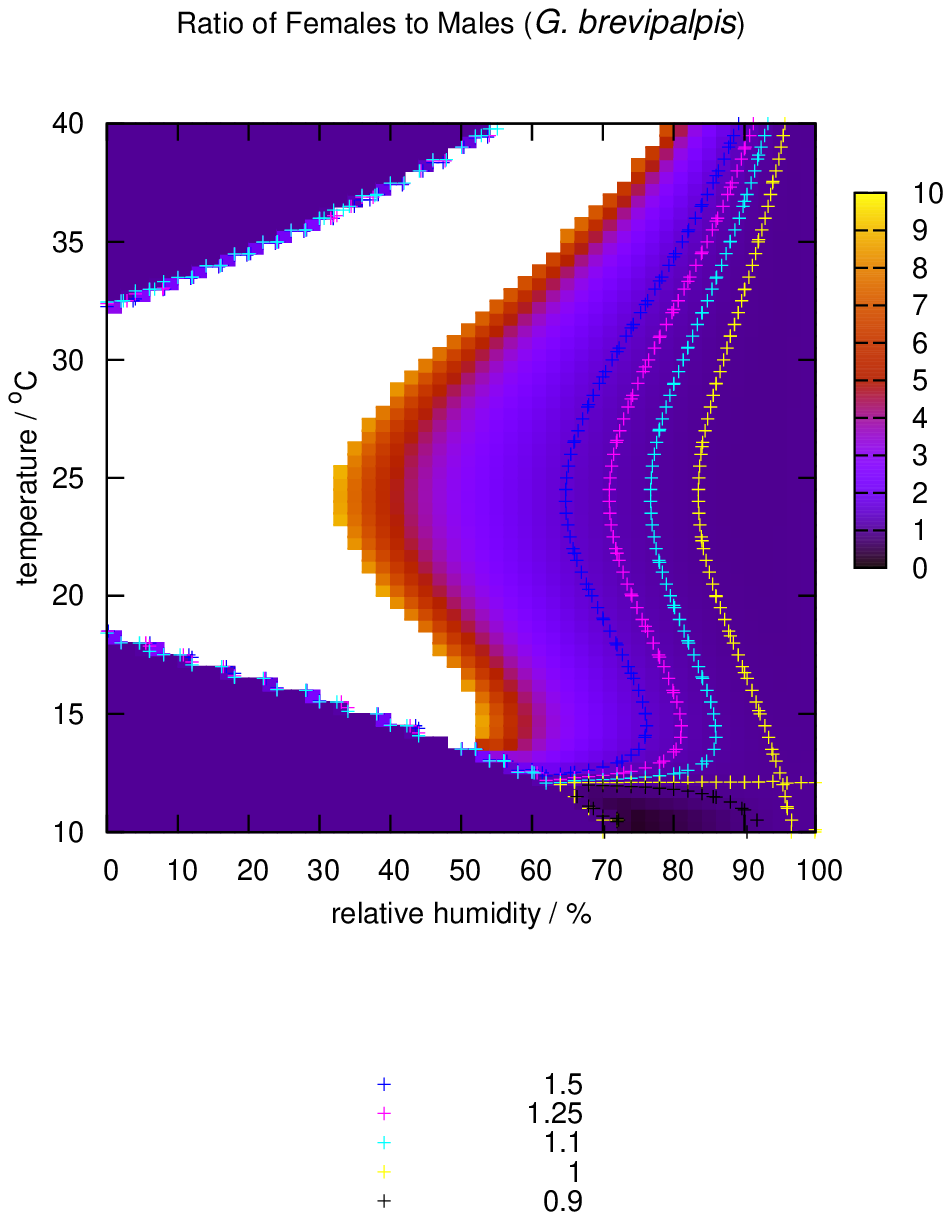}
\caption{Computed ratio of female to male emergent for {\em G. morsitans} (left) and {\em G. brevipalpis} (right). Note that this is a minimum scenario, owing to the lack of availability of any gender based data on pupal masses and surface areas (these results are based on puparial duration alone).} \label{sexRatios}
   \end{center}
\end{figure} 
Most of the experimental work needed for a model of early stage mortality has
long been complete. The main causes could be summed up in terms of dehydration,
fat loss, predation and parasitism. {\sc Bursell} \cite{Bursell2} would appear
to have a small amount of data outstanding so far as teneral water loss is
concerned; namely its dependence on temperature (for some, fixed level of
activity and humidity). Some data on relative species surface areas and loss
rates for tenerals might also be useful. The relationship between pupal fat loss
and temperature has been extensively studied by {\sc Bursell} \cite{Bursell3}
and, subsequently, {\sc Phelps} \cite{Phelps1}. A cursory inspection of that
work suggests that a few data points pertaining to teneral fat consumption's
dependence on activity (at either fixed or variable temperature) are required.
Quantitative work linking predation and parasitism to the density at pupal sites
has been carried out by {\sc Rogers} and {\sc Randolph}
\cite{RogersAndRandolph1}, although this topic can almost certainly be predicted
to require some stochastic treatment. 

Once a model of early stage mortality is completed, things might become a lot
simpler. Adult mortality is an order of magnitude lower ({\sc Hargrove}
\cite{Hargrove1}) and is thought to be trivial, likely a simple dependence on
temperature (population density becoming relevant at its higher levels).
Formulae for the first and subsequent interlarval periods are known ({\sc
Hargrove} \cite{Hargrove3}). The model which results will enable anything, from
the effect of toxins, to a population's response to changes in (or unusual)
environmental conditions, to be computed. Such a model should be able to be
interrogated for anything, from sex ratios to age structure.

Although the envisaged model would be more concerned with population dynamics
than steady-state, habitat assessment, the question of habitat assessment and
remote sensing is a topic of intense interest to entomologists and
parasitologists. It would, indeed, be of interest to know how variables, such as
rainfall-duration, NDVI and elevation, might translate into soil
humidity; alternatively, into rot holes, forest-litter and mulch, composts with
their own micro-climate of heat and moisture. Such a study might not be too
difficult to carry out and would provide an interface to weld the `top-down',
empirical approach (e.g. of {\sc Rogers} and {\sc Robinson}
\cite{RogersAndRobinson}) to the `bottom-up', fundamental approach of the model
envisaged here. Any initial disagreement between the two is likely to provide
new and profound, biological insights. For example, it is hard to imagine using
a variable anything other than NDVI to quantify both the availability and
advantage afforded by a high, shady perch, to a species at the upper limits of
physically-permitted weight. Yet drainage from remote catchment areas might
leave no other trace.

\section{Acknowledgements} 

Abdalla Latif and the Onderstepoort Veterinary Institute are thanked for their
generousity in co-funding this work. All references used in this work were the
personal property of John Hargrove. Glyn Vale and Andrew Parker are thanked for
their general advice, helpfulness and enthusiasm. The author wishes to
acknowledge the reviewers in referring him to the remote sensing work which
supports the conclusion. The usual friends, Neil Muller and Kevin Colville, are
thanked for assistance in the way of technical support.

Although obviously not the original intention, this work also turns out to be something of a tribute to the late E. Bursell. Bursell's investigation was both comprehensive and enquiring. He asked the right questions and for this we sing his praises.

\bibliography{pupalH2Oloss}

\begin{thebibliography}{10}

\bibitem{Bursell1}
E.~Bursell.
\newblock The water balance of tsetse pupae.
\newblock {\em Philosophical Transactions of the Royal Society of London},
  241(B):179--210, 1958.

\bibitem{Bursell2}
E.~Bursell.
\newblock The water balance of tsetse flies.
\newblock {\em Transactions of the Royal Entomological Society London},
  111:205--235, 1959.

\bibitem{Bursell3}
E.~Bursell.
\newblock The effect of temperature on the consumption of fat during pupal
  develpment in glossina.
\newblock {\em Bulletin of Entomological Research}, 51(3):583--598, 1960.

\bibitem{BuxtonAndLewis1}
P.~A. Buxton and D.~J. Lewis.
\newblock Climate and tsetse flies: laboratory studies upon {{\em {G}lossina
  submorsitans}} and {{\em tachinoides}}.
\newblock {\em Philosophical Transactions}, 224(B):175--240, 1934.

\bibitem{Glasgow1}
J.~P. Glasgow.
\newblock {\em The Distribution and Abundance of Tsetse}.
\newblock International Series of Monographs on Pure and Applied Biology.
  Pergamon Press, 1963.

\bibitem{Hargrove1}
J.~W. Hargrove.
\newblock Age--dependent changes in the probabilities of survival and capture
  of the tsetse, {{\em Glossina {M}orsitans {M}orsitans {W}estwood}}.
\newblock {\em Insect Science and its Applications}, 11(3):323--330, 1990.

\bibitem{Hargrove3}
J.~W. Hargrove.
\newblock {\em The Trypanosomiases}.
\newblock Editors: I. Maudlin, P. H. Holmes and P. H. Miles. {CABI} publishing,
  {O}xford, U.K., 2004.

\bibitem{HargroveAndWilliams}
J.~W. Hargrove and B.~G. Williams.
\newblock Optimized simulation as an aid to modelling, with an application to
  the study of tsetse flies, {{\em Glossina morsitans morsitans ({D}iptera:
  Glossinidae)}}.
\newblock {\em Bulletin of Entomological Research}, 88:425--435, 1998.

\bibitem{Parker1}
A.~Parker.
\newblock {\em By communication}.
\newblock 2008.

\bibitem{Phelps1}
R.~J. Phelps.
\newblock The effect of temperature on fat consumption during the puparial
  stages of {{\em Glossina morsitans morsitans Westw. (Dipt. Glossinidae)}}
  under laboratory conditions, and its implication in the field.
\newblock {\em Bulletin of Entomological Research}, 62:423--438, 1973.

\bibitem{phelpsAndBurrows1}
R.~J. Phelps and P.~M. Burrows.
\newblock Puparial duration in {{\em {G}lossina morsitans orientalis}} under
  conditions of constant temperature.
\newblock {\em Entomologia Experimentalis et Applicata}, 12:33--43, 1969.

\bibitem{RogersAndRobinson}
D.~J. Rogers and T.~P. Robinson.
\newblock {\em The Trypanosomiases}.
\newblock Editors: I. Maudlin, P. H. Holmes and P. H. Miles. {CABI} publishing,
  {O}xford, U.K., 2004.

\bibitem{RogersAndRandolph1}
David~J. Rogers and Sarah~E. Randolph.
\newblock Estimation of rates of predation on tsetse.
\newblock {\em Medical and Veterinary Entomology}, 4:195--204, 1990.

\bibitem{Solano1}
P.~Solano.
\newblock {\em By communication}.
\newblock 2008.

\end{thebibliography}

\newpage

\section*{Addendum}

\subsection{The Issue of Sub-Standard Laboratory Pupae} \label{subStandardPupae}

In {\sc Bursell} \cite{Bursell1} it is somewhat heuristically argued that the laboratory pupae in question were too small and that all emergence curves should therefore be displaced 10\% to the left (the right in those graphs). An alternative argument can also be entertained: To bring pupae that are 10\%
too small up to size would mean multiplying volume by $\frac{100}{90}$, or each
dimension by $\sqrt[3]{\frac{100}{90}}$. It would not be unreasonable to assume reserves are volume dependent whereas water loss is dependent on
surface area. 

Suppose $\$$ is the initial starting reserve of the pupa and that $g(T,t)$ is some complicated loss rate function. If emergence is dependent on some critical water loss,
\begin{eqnarray} \label{24}	
\$ - g(T,t) \ s \ ( 100 - h ) &=& 0, 
\end{eqnarray}	
for the larger, wild puparium, one then has
\begin{eqnarray} \label{25}	
\frac{100}{90} \ \$ - g(T,t) \ \left( \frac{100}{90} \right)^{\frac{2}{3}} s \ ( 100 - H ) &=& 0,
\end{eqnarray}	
where $H$ is the humidity pertaining to a larger, wild puparium. If equation \ref{24} was satisfied for $h$, then equation \ref{25} will be satisfied for the new humidity, $H$, when 	
\begin{eqnarray} \label{26}			
&& ( 100 - H ) \ = \ \sqrt[3]{\frac{100}{90}} \ ( 100 - h ) \Rightarrow H \ = \ \sqrt[3]{\frac{100}{90}} \ h - 3.57. \nonumber
\end{eqnarray}	

Thus, replacing the argument in the old emergence relation, $E(h)$, with 
\begin{eqnarray} \label{28}						
h &=& \frac{ ( H + 3.57 ) }{ \sqrt[3]{\frac{100}{90}} } \nonumber
\end{eqnarray}	
is as good as the new one, 
\begin{eqnarray} \label{29}						
E \left( \frac{H + 3.57}{\sqrt[3]{\frac{100}{90}}} \right), \nonumber
\end{eqnarray}		
so far as puparial transpiration is concerned. Although puparial transpiration rates are of the order of ten times bigger than pupal rates, the sensu strictu pupal period is usually of the order of ten times as long. In other words, the pupal phase is just as relevant. the same reasoning as above can be applied to the historical dependence (the coefficients for equation \ref{5} suggest that historical effects are as important as the prevailing humidity). Dependence on $H$ is quadratic.

Of course, observed emergence is not linearly dependent on humidity, as the
relationships depicted in Figure \ref{allSpeciesTogether} show. If, however, one
is talking about the relevance of size to straightforward water loss (and water
loss being the predominant determinant of emergence), not variation within the
species, one would question a simple, sideways shift of 10\%.	

\subsection{Does Generalizing a {\em G. morsitans}-Based Model to Other Species Work?}

On the face of it, Assumption \ref{assumption3} is certainly the most tenuous. How valid is it? Does such a simplistic approach work? Both the examples which follow suggest that the model's general puparial rate is slightly on the low side for the first few hours, something we already know to be true. 

\subsubsection*{Example 1}

Consider the conversion of first day {\em G. morsitans} transpirational values to {\em G. brevipalpis}. Then
\begin{eqnarray*}
\frac{dk}{dt}(h,T) \ = \ \phi_{\mbox{\scriptsize puparium}}(h) \ \theta_{\mbox{\scriptsize puparium}}(T) \ \delta_{\mbox{\scriptsize puparium}} \ = \ 1.10 \times 10^{-3} \times 10.3 \ = \ 11.4 \times 10^{-3} {\mbox h}^{-1}.
\end{eqnarray*}
This answer is, of course, in dimensionless, {\em G. morsitans} pupal masses. To convert to a fraction of {\em G. brevipalpis} pupal mass h$^{-1}$:
\begin{eqnarray*}
11.4 \times 10^{-3} {\mbox h}^{-1} \times \frac{ m_{\mbox{\scriptsize morsitans}} }{ m_{\mbox{\scriptsize brevipalpis}} } \ = \  11.4 \times 10^{-3} {\mbox h}^{-1} \times \frac{ 31 }{ 78 } \ = \  4.51 \times 10^{-3} {\mbox h}^{-1} 
\end{eqnarray*}
Compare this with the measured value of $4.79 \times 10^{-3} \pm 0.05 \times 10^{-3}$ pupal masses h$^{-1}$ cited in {\sc Bursell} \cite{Bursell1}. This constitutes a 6\% error.

\subsubsection*{Example 2}
Consider the conversion of first day {\em G. morsitans} transpirational values to {\em G. palpalis}. Then
\begin{eqnarray*}
\frac{dk}{dt}(h,T) \ = \ \phi_{\mbox{\scriptsize puparium}}(h) \ \theta_{\mbox{\scriptsize puparium}}(T) \ \delta_{\mbox{\scriptsize puparium}} \ = \ 1.10 \times 10^{-3} \times 2.54 \ = \ 2.80 \times 10^{-3} {\mbox h}^{-1}. 
\end{eqnarray*}
This answer, again, is in dimensionless, {\em G. morsitans} pupal masses. To convert to a fraction of {\em G. palpalis} pupal mass h$^{-1}$:
\begin{eqnarray*}
2.80 \times 10^{-3} {\mbox h}^{-1} \times \frac{ m_{\mbox{\scriptsize morsitans}} }{ m_{\mbox{\scriptsize palpalis}} } \ = \  2.80 \times 10^{-3} {\mbox h}^{-1} \times  \frac{ 31 }{ 32 } \ = \  2.70 \times 10^{-3} {\mbox h}^{-1} 
\end{eqnarray*}
Compare this with the measured value of $3 \times 10^{-3} \pm 0.4 \times 10^{-3}$ pupal masses h$^{-1}$ cited in {\sc Bursell} \cite{Bursell1}. The 10\% error is within the measured limits and tolerable by `engineering standards'.

\end{document}